%% file: main.tex
\documentclass[letterpaper,romanappendices]{IEEEtran}

\input{preamble}

\newcommand{\RCO}{R_{\op{CO}}}
\newcommand{\sRCO}{\rsfsR_{\op{CO}}}  

\newcommand{\CS}{C_{\op{S}}}
\newcommand{\RS}{R_{\op{S}}}

\newcommand{\JGK}{J_{\op{GK}}}

\title{Secret key agreement for hypergraphical sources with limited total discussion}

\author{Chung Chan
\thanks{Chung Chan (email: chung.chan@cityu.edu.hk) is with the Department of Computer Science, City University of Hong Kong.}
\thanks{The proofs of the main results were completed on Mar~22nd, 2019.}}

\begin{document}

\IEEEoverridecommandlockouts
\maketitle

\begin{abstract}
  This work considers the problem of multiterminal secret key agreement by limited total public discussion under the hypergraphical source model. The secrecy capacity as a function of the total discussion rate is completely characterized by a polynomial-time computable linear program. Compared to the existing solution for a particular hypergraphical source model called the pairwise independent network (PIN) model, the current result is a non-trivial extension as it applies to a strictly larger class of sources and a more general scenario involving helpers and wiretapper's side information. In particular, while the existing solution by tree-packing can be strictly suboptimal for the PIN model with helpers and the hypergraphical source model in general, we can show that decremental secret key agreement and linear network coding is optimal, resolving a previous conjecture in the affirmative. The converse is established by a single-letter upper bound on the secrecy capacity for discrete memoryless multiple sources and individual discussion rate constraints. The minimax optimization involved in the bound can be relaxed to give the best existing upper bounds on secrecy capacities such as the lamination bounds for hypergraphical sources, helper-set bound for general sources, the bound at asymptotically zero discussion rate via the multivariate G\'ac--K\"orner common information, and the lower bound on communication complexity via a multivariate extension of the Wyner common information.  These reductions unify existing bounding techniques and reveal surprising connections between seemingly different information-theoretic notions. Further challenges are posed in this work along with a simple example of finite linear source where the current converse techniques fail even though the proposed achieving scheme remains optimal.
\end{abstract} 


\section{Introduction}
\label{sec:intro}

The problem of secret key agreement by public discussion was formulated by \cite{ahlswede93,maurer93} where two users with correlated private observations discuss in public at unlimited rate to agree on a common secret key. The key has to be secured from a wiretapper who listens to the entire public discussion and observes some side information correlated with the users' private observations. The fact that public discussion helps generate more secret key bits was discovered in~\cite{bennett1988privacy}. A natural question is: What is the maximum secret key rate, called the secrecy capacity, achievable by a given public discussion rate? Equivalently, what is the minimum discussion rate, called the communication complexity, required to achieve a given secret key rate? Rate-limited public discussion was first considered in \cite{csiszar00}, which also introduced a helper who can help generate the secret key but needs not recover it. \cite{csiszar04} extended the problem to the multiterminal case involving arbitrary numbers of users and helpers, who can be trusted or untrusted. 

The secrecy capacity or communication complexity is, in general, unknown except in some special cases. For the two-user case, the capacity was characterized in \cite{ahlswede93} with unlimited one-way public discussion. With two-way interactive discussion but no wiretapper's side information, the capacity reduces to Shannon's mutual of the user's private observations~\cite{ahlswede93,maurer93}. If, in addition, that the number of rounds of interactive discussion is limited, the communication complexity that achieves the largest possible secrecy capacity was given by \cite{tyagi13}. Subsequently, the secrecy capacity as a function of the individual discussion rates of the two users was derived in \cite{LCV16}. If the number of rounds is unlimited, however, these characterizations are not considered single-letter solutions as they involve optimizations over an unbounded number of auxiliary random variables, which are incomputable. 

Despite the difficulty in getting a single-letter characterization of the secrecy capacity in the two-user case, the multiterminal case has been considered and resolved partially. For instance, \cite{chan18isit,chan19cs0} characterized the capacity for the case with no discussion or at asymptotically $0$ discussion rate. In the other extreme where the discussion is unlimited, the secrecy capacity was characterized in \cite{csiszar04} when there is no wiretapper's side information. The capacity as a function of the total discussion rate was characterized in \cite{chan19lamination} for the case without helpers and for a graphical source model called the pairwise independent network (PIN) model proposed in \cite{nitinawarat10}. The capacity as a function of individual discussion rate is also characterized in \cite{chan17isit} for the tree PIN model, and in \cite{zhou2018secrecy,zhou2018isit} for minimally connected hypergraphical sources. Other than the above cases, there are various bounds on the secrecy capacity and communication complexity~\cite{MKS16,chan16isit,mukherjee16,chan16itw,chan17isit,chan17oo,chan19plska,chan17cska} for general and special source models, but exact characterizations remain unknown. 

In particular, there appears no direct extension of existing solutions to cover the PIN model with helpers and hypergraphical sources that are not necessarily minimally connected. The achieving scheme for PIN model without helpers in \cite{chan17isit,chan19lamination} used the tree-packing scheme of~\cite{nitinawarat-ye10,nitinawarat10}, which is strictly suboptimal in achieving the secrecy capacity, even in the case with unlimited public discussion. The solution for minimally connected hypergraphical sources in \cite{zhou2018isit,zhou2018secrecy} was derived using a bound in \cite{chan17isit} that can be loose when the PIN model has cycles.

The focus of this work is primarily on the multiterminal setting with an arbitrary number of users and helpers. The goal is to unify different bounding techniques in existing works and improve them to give exact single-letter capacity characterizations for a larger class of source models beyond the PIN and minimally connected hypergraphical sources. In particular, decremental secret key agreement was conjectured to be optimal for hypergraphical sources~\cite{chan17isit,mukherjee16}. The conjecture can be further generalized to the optimality of compressed secret key agreement for finite linear sources~\cite{chan17cska}.  With the same linearity arguments as in \cite{chan19plska}, the conjecture implies the optimality of the linear network coding scheme~\cite{chan10phd,chan11itw,chan11delay,chan11isit,chan12ud} and that private randomization is not needed. 

In this work, we have unified and improved existing bounding techniques to show that decremental secret key agreement is optimal for hypergraphical source. Furthermore, the secrecy capacity can communication complexity for hypergraphical sources are characterized by linear programs that are polynomial-time computable. For finite linear sources, the conjecture remains unsolved, but we give an example to illustrate the limitation of the converse result and the potential improvement.
We remark that \cite{courtade16} also characterized the communication complexity for hypergraphical sources but under the assumption that the discussion is linear. Furthermore, \cite{courtade16} considered a one-shot model where the communication complexity was shown to be NP-hard to compute. In contrast, we consider an asymptotic model without assuming the discussion is linear and obtained polynomial-time computable characterizations. We also remark that there is a duality between the secret key agreement problem and the problem of generating maximum common randomness (distributed simulation)~\cite[Theorem~2.3]{chan10phd}. Hence, the results for secret key agreement can translate to the results for maximum common randomness and vice versa.

The paper is organized as follows. We formulate the secret key agreement problem for general sources and hypergraphic sources in Section~\ref{sec:problem}. Section~\ref{sec:problem} gives the main results, namely an improved converse for general sources and its reduction to the exact linear programming characterizations for hypergraphical sources. Section~\ref{sec:general} explains how the converse reduces to various existing bounds on secrecy capacity and communication complexity. Section~\ref{sec:cg} gives an example of a finite linear source for which the converse is loose. The proofs are given in the appendix.

\section{Problem formulation}
\label{sec:problem}


We consider the \emph{multiterminal secret key agreement} problem formulated in \cite{csiszar04} for a multiterminal discrete memoryless source
\begin{align*}
    \RZ_V:=(\RZ_i\mid i\in V) \text{ and } \tRZ
\end{align*}
distributed according to a given joint distribution $P_{\RZ_V\tRZ}$ over a possibly unbounded support set $Z_V\times \tilde{Z}$. $V$ is a finite set of users, $\RZ_i$ is the private source of user~$i\in V$, and $\tRZ$ is the wiretapper's side information. 

For secret key agreement, every user~$i\in V$ observes an $n$ i.i.d.\ sequence $\RZ_i^n:=(\RZ_{i1},\dots,\RZ_{in})$ and generates a private and possibly continuous random variable $\RU_i$ independent of the source, i.e.,
\begin{align}
    P_{\RU_V|\RZ_V^n} = \prod_{i\in V} P_{\RU_i}.
    \label{eq:U}
\end{align}
Then, the users engage in an \emph{interactive public discussion} where some user $i_j\in V$ at round $j\in \Set{1,\dots,r}$ of the discussion reveals in public a finitely valued message
\begin{subequations}
\label{eq:discussion}
\begin{align}
    \tRF_j := \tilde{f}_j(\RU_{i_j},\RZ_{i_j}^n,\tRF^{j-1}),\label{eq:tF}
\end{align}
namely a deterministic function of the accumulated knowledge of user~$i_j$, which includes the private randomization $\RU_{i_j}$, source $\RZ_{i_j}^n$, and all the previous discussion $\tRF^{j-1}$. For notational simplicity, we define
\begin{align}
    \RF_i&:=(\tRF_j\mid j\in \Set{1,\dots,r},i_j = i)\label{eq:Fi}\\
    \RF &:= \RF_V,\label{eq:F}
\end{align}
\end{subequations}
namely the entire discussion by user $i$ and by all users respectively. 

After the public discussion, a subset $A\subseteq V$ of the users, called the \emph{active users}, wants to agree on a secret key $\RK$ secured against a wiretapper observing $\tRZ^n$, the entire discussion $\RF$, and the source $\RZ_D^n$ of a subset $D\subseteq V`/A$ of users called the \emph{untrusted helpers}. ($V`/A$ is called the set of \emph{helpers}, where $V`/A`/D$ is the set of \emph{trusted helpers}.) More precisely, a sequence in $n$ of $\RU_V$, $\RF$ and $\RK$ is a secret key agreement scheme if there exist decoding functions $`f_i$ for $i\in A$ such that
\begin{align} 
    \lim_{n\to `8} \Pr\Set{\RK=`f_i(\RU_i,\RZ_i^n,\RF)\;\forall i\in A} &=1 \label{eq:recover}\\
    \lim_{n\to `8} \log \abs{K} - H(\RK|\RF,\RZ_D^n,\tRZ^n) &= 0\label{eq:secrecy},
\end{align}
where $K$ denotes a finite set of possible secret key values. The prior condition~\eqref{eq:recover} is called the \emph{recoverability constraint} and the latter one~\eqref{eq:secrecy} is called the \emph{secrecy (and uniformity) constraint}. 

A secret key rate $r_{\opK}\in `R_+$ is said to be achievable by the public discussion rates $r_V:=(r_i\mid i\in V)\in `R_+^V$ if and only if there exists a secret key agreement scheme $(\RK,\RF)$ satisfying the above conditions \eqref{eq:recover} and \eqref{eq:secrecy}, and the rate constraints
\begin{subequations}
\begin{align}
    r_{\opK} &\leq \liminf_{n\to `8} \frac1n \log\abs{K}\label{eq:rK}\\
    r_i &\geq \limsup_{n\to `8} \frac1n \log \abs{F_i} \quad\forall i\in V,\label{eq:ri}
\end{align}
\end{subequations}
where $F_i$ denotes the set of possible messages by user $i$.

Denotes the set of achievable rate tuple $(r_{\opK},r_V)$ by $\rsfsR^{V,A,D}[\RZ_V\|\tRZ]$. The secrecy capacity is the maximum achievable key rate denoted by
\begin{align}
\begin{split}
    \CS^{V,A,D}[\RZ_V\|\tRZ](r_V) &:= \sup \{r_{\opK}\in `R_+\mid\\ &\kern-2em (r_{\opK},r_V)\in \rsfsR^{V,A,D}[\RZ_V\|\tRZ]\},\end{split}\label{eq:CSrV}\\
    \begin{split}
    \CS^{V,A,D}[\RZ_V\|\tRZ](R) &:= \sup \Set{r_{\opK}\in `R_+\mid\\ &\kern-2em (r_{\opK},r_V)\in \rsfsR^{V,A,D}[\RZ_V\|\tRZ], r(V)\leq R},
    \end{split}\label{eq:CSR}
\end{align}
for $ r_V\in `R^V_+$ and $R\in `R_+$,
where, for notational convenience, we define for $B\subseteq V$
\begin{align*}
r(B):=\sum_{i\in B} r_i.
\end{align*}

It is easy to see that the secrecy capacity is non-decreasing in the discussion rates. The communication complexity as a function of the key rate $r_{\opK}$ is defined as
\begin{align}
\begin{split}
    \RS^{V,A,D}[\RZ_V\|\tRZ](r_{\opK}) &:= \inf \{R\in `R_+\mid\\ &\kern-4em \CS^{V,A,D}[\RZ_V\|\tRZ](R)=\CS^{V,A,D}[\RZ_V\|\tRZ](`8)\},\end{split}\label{eq:RSrK}
\end{align}
which is simply the inverse of $\CS^{V,A,D}[\RZ_V\|\tRZ](R)$. In the rest of the paper, we will omit the dependency on $V,A,D,\RZ_V,\tRZ$ and write
\[\rsfsR, \CS(r_V), \CS(R), \RS(r_{\opK})\]
for the case without untrusted helper ($D=`0$) nor wiretapper's side information ($\tRZ=0$) unless otherwise specified. We will also write 
\begin{align}
  \CS&=\liminf_{R\to`8}\CS(R)\label{eq:CS}\\
  \RS&=\limsup_{r_{\opK}\to \CS}\RS(r_{\opK})\label{eq:RS}
\end{align}
respectively for the \emph{unconstrained secrecy capacity} with unbounded discussion rate, and the communication complexity to attain the unconstrained the secrecy capacity. 
%

Consider a hypergraph $(V,E,`x)$ with vertex set $V$, edge set $E$ and the edge function $`x:E\to 2^{V}`/\Set{`0}$, where $`x(e)$ denotes the set of vertices incident on an edge $e\in E$. A hypergraphical source is defined with respect to such a hypergraph as~\cite{chan10phd,chan10md}
\begin{subequations}
\label{eq:hyp}
\begin{align}
  \RZ_i &= (\RX_e \mid i\in `x(e)) && \text{for $i\in V$}\label{eq:hyp:Z}
\end{align}
where $\RX_E:=(\RX_e\mid e\in E)$ is a given random vector and $\RX_e$'s are independent random variables called the edge (random) variables with bounded supports. 
In other words, each user gets to observe a subset of the independent edge variables. The model covers the pairwise independent network (PIN) model~\cite{nitinawarat10,nitinawarat-ye10} as a special case when exactly two users observe each edge variable, i.e., $\abs{`x(e)}=2$, and so the hypergraph reduces to a graph. 
One may also consider a hypergraphical source model with wiretapper's side information, where
\begin{align}
    \tRZ = \RX_{\tdE}\label{eq:hyp:tRZ}
\end{align}
\end{subequations}
for some given subset $\tdE\subseteq E$.

\section{Main results}
\label{eq:results}

All the converse results in this paper are based on the following single-letter upper bound on the secrecy capacity. The bound applies to general sources and allows for trusted helpers, i.e., with $D=`0$ but $V`/A$ possibly non-empty.

\begin{Theorem}
  \label{thm:general}
  The secrecy capacity~\eqref{eq:CSrV} as a function of discussion rate $r_V\in `R^V$ is upper bounded as follows:
  \begin{subequations}
    \label{eq:CSrV:ub}
    \begin{align}
      \CS(r_V) &\leq \sup_{\RW,r'_V} I(\RW\wedge \RZ_V) - r'(V) \label{eq:CSrV:ub1} \\
      &= \sup_{\RW} \min_{`l\in \bar{`L}} I(\RW \wedge \RZ_V) -r(V) \notag\\
      &\kern1em -\sum_{B\nsupseteq A} `l(B)`1[I(\RW\wedge \RZ_B|\RZ_{V`/B}) - r(B)`2],\label{eq:CSrV:ub2} 
    \end{align}
  \end{subequations}
  where 
  \begin{itemize}
    \item the maximization in the first expression~\eqref{eq:CSrV:ub1} is over the choices of an auxiliary random variable $\RW$ (or its distribution $P_{\RW|\RZ_V}$ more specifically) and a real vector $r'_V\in `R^V$ subject to the constraints
    \begin{subequations}
      \label{eq:r'_V}
      \begin{align}
        r'(B) &\geq I(\RW\wedge \RZ_B|\RZ_{V`/B}) && \forall B\nsupseteq A\label{eq:r'_V1}\\
        r'_i &\leq r_i && \forall i\in V;\label{eq:r'_V2}
      \end{align}
    \end{subequations} 
    \item the minimization in the second expression~\eqref{eq:CSrV:ub2} is over the choice of a set function $`l:\Set{B\subseteq V\mid B\nsupseteq A}\to `R_+$ satisfying 
    \begin{align}
      \sum_{B\nsupseteq A:i\in B} `l(B) &\geq 1 && \forall i\in V.\label{eq:fc}
    \end{align}
    $`l$ is referred to as a \emph{fractional cover} and we denote the set of all such fractional covers by $\bar{`L}$.
  \end{itemize}
\end{Theorem}

\begin{Proof}
  See Appendix~\ref{sec:proof:thm:general}.
\end{Proof}

An important simplification is to consider the total discussion rate constraint instead of the individual rates. The above bound translates directly to the following bound on the secrecy capacity for a given total discussion rate.
\begin{Corollary}
  \label{cor:general}
  The secrecy capacity~\eqref{eq:CSR} for $R\geq 0$ are upper bounded as follows:
  \begin{subequations}
    \label{eq:CSR:ub}  
    \begin{align}
      \CS(R) 
      &\leq \sup
      \{ I(\RW\wedge \RZ_V) - r(V) \mid r(V)\leq R,\notag\\
      &\kern3em r(B) \geq I(\RW\wedge \RZ_B|\RZ_{V`/B}) \quad \forall B\nsupseteq A
      \}\label{eq:CSR:ub1}\\
      &= \sup_{\RW:`r(\RW)\leq R} I(\RW\wedge \RZ_V) - `r(\RW) \label{eq:CSR:ub2}
    \end{align}
  \end{subequations}
  \begin{align}
    \CS &\leq \min_{`l\in `L} I_{`l}(\RZ_V),\kern11em \label{eq:CS:ub}
  \end{align}
  where
  \begin{align}
    `r(\RW) &:= \max_{`l\in `L} \sum_{B\nsupseteq A} `l(B) I(\RW\wedge \RZ_B|\RZ_{V`/B})\label{eq:rho}\\
    I_{`l}(\RZ_V) &:= H(\RZ_V) - \sum_{B\nsupseteq A}`l(B)H(\RZ_B|\RZ_{V`/B})\label{eq:I_l}
  \end{align}
  and $`L$ denotes the set of fractional partitions $`l:\Set{B\subseteq V\mid B\nsupseteq A}\to `R_+$, which are fractional covers with the constraint~\eqref{eq:fc} satisfied with equality, i.e.,
  \begin{align}
    \sum_{B\nsupseteq A:i\in B} `l(B) &= 1 && \forall i\in V.\label{eq:fp}
  \end{align}
  The corresponding lower bound on the communication complexity  \eqref{eq:RSrK},
  \begin{align}
    \RS(r_{\opK}) &\geq \inf_{\RW:I(\RW\wedge \RZ_V)-`r(\RW)\geq r_{\opK}\kern-2em} \kern-2em \raisebox{2pt}{$I(\RW\wedge \RZ_V)-r_{\opK} $}\label{eq:RSrK:lb}
  \end{align}
  for any given secret key rate $r_{\opK}\geq 0$.
\end{Corollary}

\begin{Proof}
  See Appendix~\ref{sec:proof:cor:general}.
\end{Proof}

Note that the above bounds involve an auxiliary random variable $\RW$, an optimal solution of which exists by standard support lemma~\cite{csiszar81} if the support of the random source is finite. It is also straightforward to argue that the upper bounds on the secrecy capacity are non-decreasing, concave, and continuous in the discussion rates. 
The two expressions (\eqref{eq:CSrV:ub1} and \eqref{eq:CSrV:ub2}) for the upper bound~\eqref{eq:CSrV:ub} are related by the linear programming duality~\cite{schrijver02}, where the minimization over $`l$ is the dual of the maximization over $r'_V$. $`l(B)$ is the Lagrangian multiplier for the constraints on $r'_V$ in \eqref{eq:r'_V}. 

It is instructive to compare $r'_V$ to the feasible rate of communication for omniscience~\cite{csiszar04} in
\begin{subequations}
  \label{eq:sRCO}
  \begin{align}
    \sRCO &= \Set{r_V\in `R^V\mid \\
    &\kern1em r(B)\geq H(\RZ_B|\RZ_{V`/B}) \kern1em \forall B\subseteq V:B\nsupseteq A}.\label{eq:SW}
  \end{align}
\end{subequations}
The above corresponds to the set of public discussion rate tuple such that each user can recover the entire source $\RZ_V$ after the discussion, i.e., attain \emph{omniscience}. Suppose the source $\RZ_V$ has finite support. Then, the region must be non-empty. The constraints in~\eqref{eq:r'_V1} play a similar role as the Slepian-Wolf constraints in~\eqref{eq:SW} above. In particular, the two sets of constraints are the same if $\RW=\RZ_V$ and $r'_i=r_i$. This connection can be observed similarly in the bounds for total discussion rate instead of individual discussion rate. In particular, with $\RW=\RZ_V$, $`r(\RW)$ defined in \eqref{eq:rho} becomes the smallest rate of communication for omniscience~\cite{csiszar04}
\begin{subequations}
  \label{eq:RCO}
  \begin{align}
    \RCO &= \min_{r_V\in \sRCO} r(V)\\
    &= \max_{`l\in `L} \sum_{B\nsupseteq A} `l(B) H(\RZ_B|\RZ_{V`/B})
  \end{align}
\end{subequations}
where the last equality is again by the linear programming duality. With $`r(\RW)=\RCO$ and assuming $\RZ_V$ has finite support, the bound~\eqref{eq:CSR:ub2} on secrecy capacity becomes \eqref{eq:CS:ub}, which is the unconstrained secrecy capacity characterized in \cite{csiszar04} as
\begin{subequations}
  \label{eq:CSRCO}
  \begin{align}
    \CS &= H(\RZ_V) - \RCO\label{eq:CSRCO1} \\
    &= \min_{`l\in `L} I_{`l}(\RZ_V)\label{eq:CSRCO2},
  \end{align}
\end{subequations}
where $I_{`l}$ is the information measure defined in \eqref{eq:I_l}. The last equality means that the bound \eqref{eq:CS:ub} is tight when $\RZ_V$ has finite support. The expression is non-negative as expected because by the Shearer Lemma (see \cite[Lemma~D.1]{chan17ooa} or \cite{csiszar08,chan10phd})
\begin{align}
  I_{`l}(\RZ_V)\geq 0 \label{eq:shearer}
\end{align}
with equality if $\RZ_i$'s are mutually independent. In the case without helpers, i.e., $A=V$, \eqref{eq:CSRCO2} can be further simplified to the following multivariate mutual information as shown in \cite{chan2008tightness,chan10md,chan15mi}:
\begin{subequations}
  \begin{align}
    I(\RZ_V) 
    &= \min_{\mcP\in \Pi'(V)}\frac1{\abs{\mcP}} D(P_{\RZ_V}\|\prod_{C\in \mcP} P_{\RZ_C})\label{eq:MMI:1}\\
    &:= \inf\Set{`g\in `R|\forall\mcP\in \Pi'(V),\notag\\
    &\kern3em H(\RZ_V)-`g=\sum_{C\in \mcP} H(\RZ_C)-`g}\label{eq:MMI:2}
  \end{align}    
\end{subequations}
where $\Pi'(V)$ is the set of partitions of $V$ into at least two non-empty disjoint sets.
The first expression~\eqref{eq:MMI:1} was given as an upper bound on $\CS$ in \cite{csiszar04}. It can be obtained from \eqref{eq:I_l} with
\begin{align}
  `l(B) = \begin{cases}
    \frac1{\abs{\mcP}-1}, & V`/B\in \mcP\\
    0, & \text{otherwise.}
  \end{cases}
\end{align}
for any partition $\mcP$ of $V$. The constraint in the second expression~\eqref{eq:MMI:2} is the constrained \emph{residual independence relation} given in \cite{chan15mi,chan16cluster}, which means that $I(\RZ_V)$ is the smallest amount of shared information removal of which leads to independence.


The bounds on the secrecy capacity and communication complexity can be shown to be tight for the hypergraphical sources as follows:

\begin{Theorem}
    \label{thm:hyp}
      For hypergraphical sources defined in \eqref{eq:hyp:Z}, the secrecy capacity~\eqref{eq:CSR} and communication complexity~\eqref{eq:RSrK} are equal to the upper bound~\eqref{eq:CSR:ub2} and lower bound~\eqref{eq:RSrK:lb} respectively, which can be simplified further by setting 
      \begin{subequations}
        \label{eq:hyp:sc}
        \begin{align}
          \RW &=(\RQ_E,\RX'_E) && \text{where}\label{eq:hyp:W}\\
          \RX'_e &= \begin{cases}
            \RX & \RQ_e=1\\
            0 & \RQ_e=0
          \end{cases} &&\text{for $e\in E$}
          \label{eq:hyp:X'} 
        \end{align}
        and $\RQ_e$'s are independent bits independent of the source with distribution
        \begin{align}
          \overbrace{P_{\RQ_E|\RX_E}}^{P_{\RQ_E|\RZ_V}} = \prod_{e\in E} \overbrace{\op{Bern}`1(\tfrac{x_e}{H(\RX_e)}`2)}^{P_{\RQ_e}}
          \label{eq:hyp:Q}
        \end{align} 
      \end{subequations}
      for some vector $x_V\in `R^V$.
    \end{Theorem}
 
    \begin{Proof}
        See Appendix~\ref{sec:proof:thm:hyp}.
    \end{Proof}

    
    \begin{Corollary}
      \label{cor:hyp}
      For hypergraphical sources, $R,r_{\opK}\geq 0$,
      \begin{subequations}
        \label{eq:hyp:CSR}
        \begin{align}
          \begin{split}
            \CS(R) 
            &= \max\{ x(E) - r(V) | r(V)\leq R\\
              &\kern3em r(B)\geq x(E(B)) \quad\forall B\subseteq V:B\nsupseteq A\\
              &\kern3em 0\leq x_e\leq H(\RX_e) \quad \forall e\in E
            \}
          \end{split}\label{eq:hyp:CSR1}\\
          \begin{split}
            &= \max\{ x(E) - `r | `r \leq R\\
              &\kern3em`r = \max_{`l\in `L} \sum_{B\nsupseteq A} `l(B) x(E(B))\\
              &\kern3em 0\leq x_e\leq H(\RX_e) \quad \forall e\in E
            \}
          \end{split}\label{eq:hyp:CSR2}
        \end{align}
      \end{subequations}
      \begin{align}
        \begin{split}
          \RS(r_{\opK})&= \min\{ x(E) - r_{\opK} \mid \\
            &\kern3em 0\leq x_e\leq H(\RX_e) \quad \forall e\in E\\
            &`1.\kern3em  x(E) - \sum_{B\nsupseteq A}`l(B) x(E(B))\geq r_{\opK}
          `2\}
        \end{split}\label{eq:hyp:RS}
      \end{align}
      where we define $x(E'):=\sum_{e\in E'} x_e$ for $E'\subseteq E$ as usual and
      \begin{align}
        E(B)&:=\Set{e\in E\mid `x(e) \subseteq B} \quad \text{for }B\subseteq V,\label{eq:EB}
      \end{align}
      namely, the set of edges that are incident only on nodes within $B$. The linear programs above can be solved in polynomial-time. 
    \end{Corollary}
    
    \begin{Proof}
      See Appendix~\ref{sec:proof:cor:hyp}.
    \end{Proof}
    
    The theorem is proved by showing that the bounds~\eqref{eq:CSR:ub2} and \eqref{eq:RSrK:lb} on the secrecy capacity can be achieved by the \emph{decremental secret key agreement} scheme in \cite{chan16isit,chan17idska}. The linear programs in the corollary are obtained by evaluating the bounds~\eqref{eq:CSR:ub2} and \eqref{eq:RSrK:lb} explicitly with the optimal solution choice of the auxiliary random variable $\RW$ in \eqref{eq:hyp:sc}.
    
    The idea of decremental secret key agreement is to reduce the randomness of the source $\RZ_V$ by eliminating some randomness of each edge variable, leading to a reduced source $\RZ'_V$, and then generate the secret key via omniscience of the reduced source, i.e., achieving the unconstrained secrecy capacity of $\RZ'_V$. More formally, the reduced source is
    \begin{align}
      \RZ'_i &:= (\RX'_e \mid i\in `x(e)) && \text{for $i\in V$}\label{eq:hyp:Z'}
    \end{align}
    where $\RX'_e$
    and $\RQ_e$ are as defined in \eqref{eq:hyp:sc} 
    for some vector $x_V\in `R^V$. Each user~$i\in V$ can privately reduce their source $\RZ_i$ effectively to $\RZ'_i$ by keeping only the first $\frac{x_e}{H(\RX_e)}$ fraction of the $n$ i.i.d.\ samples of $\RX_e$ for each edge $e\in E$ with $i\in `x(e)$. An immediate generalization of decremental secret key agreement to general sources beyond hypergraphic sources is the \emph{compressed secret key agreement} in \cite{chan17cska}, where the reduced source $\RZ'_i$ can be chosen as arbitrary processing of $\RZ_i$ with a time sharing variable $\RQ$, i.e.,
    \begin{align}
      H(\RZ'_i|\RZ_i,\RQ)&=0\\
      P_{\RQ|\RZ_V} &= P_{\RQ}.
    \end{align}
    The secret key rate $\CS[\RZ'_V|\RQ]$ is therefore achievable by a discussion of rate $\RCO[\RZ'_V|\RQ]$, where, similar to the conditional entropy $H(\RZ'_V|\RQ)$, $\CS$ and $\RCO$ evaluated at $\RZ'_V|\RQ$ means conditioning on $\RQ$, i.e., with distribution $P_{\RZ'_V|\RQ}(\cdot|\RQ)$ as the source, and then take expectations with respect to $\RQ$.
    
    The optimality of decremental secret key agreement resolved the conjecture in \cite{mukherjee16} that decremental secret key agreement is optimal and also the conjecture in \cite{chan19lamination} that linear network coding (discussion) is optimal. The idea of secret key agreement by linear network coding can be found in \cite{chan11itw,chan11delay,chan11isit,chan12ud,chan13itw,chan13isit}. A straightforward extension of the results to the case with untrusted helpers and wiretapper's side information is as follows:
    
    \begin{Proposition}
      \label{pro:hyp:w}
      In the case with untrusted helpers ($D\neq `0$) and wiretapper's side information \eqref{eq:hyp:tRZ}, we have 
      \begin{align}
          \CS^{V,A,D}[\RZ_V\|\RW](R) = \CS^{V`/D,A,`0}[\RZ'_{V`/D}\|0](R)\quad \text{where}
          \label{eq:hyp:w}
      \end{align}
      \begin{align}
          \RZ'_i &:= (\RX_e \mid e\in E`/\tdE,\\
          &\kern3em i\in `x(e),D\cap `x(e)=`0) \quad\text{for } i\in V`/D,
          \label{eq:hyp:y}
      \end{align}
      which is obtained from $\RZ_i$ by removing the edge variables observed by the untrusted helpers and wiretappers.  
    \end{Proposition}

    \begin{Proof}
      Note that $\geq$ for \eqref{eq:hyp:w} holds because $\RZ'_i$ can be obtained from $\RZ_i$ for $i\in V$, and $\RZ'_{V`/D}$ is independent of $()\RZ_D,\tRZ)$.
      To explain the reverse inequality, note that the capacity does not decrease by turning the wiretapper into an untrusted helper, i.e.,
    \begin{align*}
      \CS^{V,A,D}[\RZ_V\|\RW](R) \leq \CS^{V\cup \Set{0},A,D\cup \Set{0}}[\RZ_{V\cup \Set{0}}](R)
    \end{align*}
    with $\RZ_0=\tRZ$, assuming $0\not\in V$  without loss of generality.
    The upper bound above can be further upper bounded by the R.H.S.\ of \eqref{eq:hyp:w} as desired because $P_{\RY_{V`/D}}=P_{\RZ_V|\RZ_D,\tRZ}$ by the independence of edge variables. 
    \end{Proof}

\section{Reduction to various converse results}
\label{sec:general}

In this section, we will show that the bounds in Theorem~\ref{thm:general} and Corollary~\ref{cor:general} unify various exiting converse results. First of all, by the result of \cite{csiszar04} that the upper bound~\eqref{eq:CS:ub} on $\CS$ can be achieved via communication for omniscience at the smallest rate, it is straightforward to show that, for any smallest omniscience rate tuple $r_V\in \sRCO:r(V)=\RCO$ and sum rate $R\geq \RCO$, the upper bounds~\eqref{eq:CSrV:ub}, \eqref{eq:CSR:ub2}, and \eqref{eq:CS:ub} are tight, equal to the unconstrained secrecy capacity given by \eqref{eq:CSRCO}. Furthermore, $\RW=\RZ_V$ is optimal to the maximizations in the upper bounds~\eqref{eq:CSrV:ub} and \eqref{eq:CSR:ub2}. 
Indeed, if $\RW=\RZ_V$ is also optimal to the minimization in the lower bound~\eqref{eq:RSrK:lb}, then the lower bound is also tight:

\begin{Proposition}
  For $\RZ_V$ with finite support,
  $\RS=\RCO$, i.e., the communication for omniscience scheme in \cite{csiszar04} for secret key agreement achieves $\RS$, if $\RW=\RZ_V$ is an optimal solution to the minimization in the lower bound~\eqref{eq:RSrK:lb} on the communication complexity for some $r_{\opK}\in [0,\CS]$. In particular, this holds if $\RZ_V$ has finite support and
  \begin{align}
    \min_{\RW: I_{`l}(\RZ_V|\RW) = 0,\forall `l\in `L^*} I(\RW\wedge \RZ_V) = H(\RZ_V) \label{eq:OO}
  \end{align}
  where $`L^*$ is the set of optimal solutions $`l$ to \eqref{eq:CS:ub}.
\end{Proposition}

\begin{Proof}
   Suppose $\RW=\RZ_V$ is optimal to \eqref{eq:RSrK:lb}. Then, \eqref{eq:RSrK:lb} becomes
   \begin{align*}
    \RS(r_{\opK}) &\geq H(\RZ_V) - r_{\opK} \\
    &\geq H(\RZ_V) - \min_{`l\in `L} I_{`l}(\RZ_V) && \text{by \eqref{eq:CS:ub}}\\
    &= \RCO && \text{by \eqref{eq:CSRCO2}.}
   \end{align*}
   Equality holds as desired since $\RS\leq \RCO$ by the omniscience scheme for secret key agreement in \cite{csiszar04}. 

   Next, we show that \eqref{eq:OO} implies $\RW=\RZ_V$ is optimal to \eqref{eq:RSrK:lb} with $r_{\opK}=\CS$ as follows. By \eqref{eq:RSrK:lb}
   \begin{align*}
     \RS(\CS) &\geq \inf_{\RW:I(\RW\wedge \RZ_V)-`r(\RW)\geq \CS \kern-2em} \kern-2em \raisebox{2pt}{$I(\RW\wedge \RZ_V)-\CS $}\\
     &\utag{a}= \inf_{\RW:\min_{`l\in `L} `1[I_{`l}(\RZ_V) - I_{`l}(\RZ_V|\RW)`2] \geq \min_{`l\in `L} I_{`l}(\RZ_V) \kern-2em} \kern-2em \raisebox{2pt}{$I(\RW\wedge \RZ_V) $}-\CS\\
     &\utag{b}\geq \inf_{\RW: I_{`l}(\RZ_V|\RW)=0,\forall `l\in `L^* \kern-2em} \kern-2em \raisebox{2pt}{$I(\RW\wedge \RZ_V) $}-\CS.
   \end{align*}
   \uref{a} is obtained by rewriting the constraint by \eqref{eq:I_l:exp} and $\CS(R)=\min_{`l\in `L} I_{`l}(\RZ_V)$ by \eqref{eq:CSRCO}, which in turn holds as $\RCO<`8$ for $\RZ_V$ with finite support. \uref{b} is because, for any $`l\in `L^*$, the constraint implies that
   \begin{align*}
    I_{`l}(\RZ_V) - I_{`l}(\RZ_V|\RW) &\geq I_{`l}(\RZ_V),\quad \text{or equivalently}\\
    I_{`l}(\RZ_V|\RW) &= 0.
   \end{align*}
   Finally, if \eqref{eq:OO} holds, we have
   \begin{align*}
    \RS(\CS) \geq H(\RZ_V) -\CS = \RCO
   \end{align*}
   as desired by \eqref{eq:CSRCO}.
\end{Proof}

The above result covers the sufficient condition in \cite{chan17oo,chan17ooa}. More precisely, the sufficient condition in \cite{chan17ooa} is in terms of the multivariate Wyner common information defined below for a fractional partition $`l$ as
\begin{align}
  C_{\opW,`l}(\RZ_V) &:= \inf_{\RW} I(\RW\wedge \RZ_V) && \text{such that}\\
  I_{`l}(\RZ_V|\RW) &= 0. 
\end{align}
\eqref{eq:OO} can be rewritten as $C_{\opW,`l}(\RZ_V)=H(\RZ_V)$, which is the sufficient condition in \cite{chan17ooa} with helpers. 

In the other extreme where the discussion rate has to be $0$, the upper bounds on the secrecy capacity are also tight, which cover the result in \cite{chan19cs0} with helpers.

\begin{Proposition}
  With $r'_V=\M0$ and $R=0$, the secrecy capacity upper bounds~\eqref{eq:CSrV:ub} and \eqref{eq:CSR:ub2} are tight and simplifies to the G\'acs--K\"orner common information
  \begin{subequations}
    \label{eq:JGK}
    \begin{align}
      \JGK(\RZ_A)&:= \max\{H(\RG)| \abs{G} <`8,\label{eq:|G|}\\
      &H(\RG|\RZ_i)=0 \quad \forall i\in A\}.\label{eq:cf}
    \end{align}
  \end{subequations}
 Furthermore, the optimal solution $\RG$, called the \emph{maximal common function} of $\RZ_i$ for $i\in A$, is an optimal solution for $\RW$ in \eqref{eq:CSrV:ub} and \eqref{eq:CSR:ub2}.
\end{Proposition}

\begin{Proof}
   With $r_V=0$, \eqref{eq:r'_V1} and \eqref{eq:r'_V2} implies that a feasible $\RW$ must satisfy
  \begin{align*}
    0 &= I(\RW \wedge \RZ_B|\RZ_{V`/B}) \quad \forall B\subseteq V:B\nsupseteq A\\
    &= I(\RW\wedge \RZ_{V`/\Set{i}}|\RZ_i) \quad \forall i\in A
  \end{align*}
  where the last equality is obtained by setting $B=V`/\Set{i}$. By the double Markov inequality~\cite[Problem 16.25]{csiszar2011information},
  \begin{align*}
    I(\RW \wedge \RZ_V|\RG) = 0
  \end{align*}
  for the optimal solution $\RG$ to \eqref{eq:JGK}. It follows that
  \begin{align*}
    I(\RW \wedge \RZ_V) &\leq I(\RW,\RG \wedge \RZ_V)\\
    &= I(\RG\wedge \RZ_V) + \underbrace{H(\RW\wedge \RZ_V|\RG)}_{=0}\\
    &= H(\RG),
  \end{align*}
  which implies by \eqref{eq:CSrV:ub1} that 
  \begin{align*}
    \CS(r_V)\leq H(\RG) = \JGK(\RZ_A).  
  \end{align*}
  Equality holds as desired as we can use the entire randomness of $\RG$ for the secret key without any discussion. More precisely, by \cite[Lemma~B.1]{csiszar04}, a key rate of $H(\RG)-r(V)=H(\RG)$ is achievable.
\end{Proof}

The constraint~\eqref{eq:|G|} requires $\RG$ to have finite support. If we set $\RW=\RG$, then $\rho(\RW)=0$ in \eqref{eq:rho} by the constraint~\eqref{eq:cf} that $\RG$ is a function of $\RZ_i$ for any $i\in A$, and hence the name common function. The secrecy capacity upper bound~\eqref{eq:CSR:ub2} then becomes $H(\RG)=\JGK(\RZ_A)$. The bound is achievable intuitively because $\RG$ is a common function of the active users and so, even with no discussion, a common secret key can be extracted from $\RG$ at rate $H(\RG)$. 

Other than the two extreme cases with unlimited or $0$ discussion rate, the secrecy capacity upper bound~\eqref{eq:CSrV:ub} strictly improves the existing bounds for multiterminal secret key agreement. In particular, it implies the following result that not only covers the bound in~\cite[Theorem~4.1]{chan17isit} for general sources but also extends it to the case with helpers.

\begin{Proposition}
  \label{pro:HS}
  We have $\CS(r_V)\geq r_{\opK}$ only if 
  \begin{align}
    r(S) \geq \frac1{\sum_{B\in \mcH} `l'(B)-1} `1[ r_{\opK} - I_{`l'}(\RZ_{V`/S}) `2]\label{eq:HS}
  \end{align}
  For all $S\subseteq V:\abs{A`/S}\geq 2$ and $`l':\Set{B\subseteq V`/S: B\nsupseteq A`/S}\to `R_+$ satisfying
  \begin{align}
    \sum_{B\nsupseteq A`/S} `l'(B)=1,\label{eq:HS:`l'}
  \end{align}
  i.e., $`l'$ is a fractional partition of $V`/S$.
\end{Proposition}

\begin{Proof}
  See Appendix~\ref{sec:proof:HS}
\end{Proof}

It is instructive to consider the condition on $r_{\opK}$ where the bound becomes trivial for a given choice of $S$, i.e., $r(S)\geq 0$. Since the factor $\frac1{\sum_{B\in \mcH} `l'(B)-1}$ is strictly positive, the bound is trivial only if
\[r_{\opK}\leq \min_{`l'} I_{`l'}(\RZ_{V`/S}).\] 
The condition is rather intuitive because, by \eqref{eq:CSRCO2}, the expression on the right is the unconstrained secrecy capacity when $S$ is removed or not allowed to discuss. \eqref{eq:HS} is called the \emph{helper-set bound} because it gives how much discussion (help) users in $S$ need so that users in $V`/S$ can share a key at a rate beyond their capacity.

Although the bound~\eqref{eq:HS} looks quite different from the original bound~\eqref{eq:CSrV:ub2}, it can be derived directly from the original bound by exchanging the maximization and minimization and then restricting the set of possible $`l$ appropriately. The bound in \cite[Theorem~4.1]{chan17isit} for the case without helpers, i.e., $A=V$, can be obtained from \eqref{eq:HS} with
\begin{align}
  `l'(B) = \begin{cases}
    \frac1{\abs{\mcP}-1}, & V`/B\in \mcP\\
    0, & \text{otherwise.}
  \end{cases}
\end{align}
for any partition $\mcP$ of $V`/S$. 

For hypergraphical sources, since the characterizations of the secrecy capacity in \eqref{eq:hyp} is tight, it covers the lamination bounds in \cite{chan19lamination}. The following result unifies the lamination bounds:
\begin{Proposition}
  \label{pro:LB}
  For hypergraphical sources and $R\geq 0$,
  \begin{align}
    0\leq \CS(R) - H(\RX_{E`/E'}) \leq `1(\frac1{\max_{`l\in `L}`a(`l)}-1`2) R\label{eq:LB}
  \end{align}
  where
  \begin{align}
    E'&:=\Set{e\in E|`x(e)\nsupseteq A} \label{eq:E'}\\
    `a(`l)&:=\begin{cases}\displaystyle
      \min_{e\in E'} \sum_{B\nsupseteq A: `x(e)\subseteq B} `l(B), & H(\RX_{E'})\neq 0\kern-1em \\
      1, & \text{otherwise.}
    \end{cases}
    \label{eq:`a}
  \end{align}
  Equality holds if $R \leq \min_{e\in E'} H(\RX_e)$.
\end{Proposition}

\begin{Proof}
  See Appendix~\ref{sec:proof:LB}.
\end{Proof}

Note that the upper bound is linear in $R$, and the slope can be bounded as follows. 
\begin{Proposition}
  \label{pro:`a}
  For hypergraphical sources, the slope of $R$ in the upper bound~\eqref{eq:LB} can be bounded as follows:
  \begin{align}
    0 &\leq \frac1{\max_{`l\in `L}`a(`l)} - 1 \leq \frac{\min\Set{d,\abs{A}}-1}{\max\Set{\abs{A}-d,1}} \quad \text{where}\label{eq:slope}\\
    d&:=\max_{e\in E'} \abs{`x(e)}\label{eq:d}
  \end{align}
  denotes the maximum degree of the edges in $E'$. Furthermore, the bounds in \eqref{eq:slope} can be achieved with equality for some hypergraphs.
\end{Proposition}

\begin{Proof}
  See Appendix~\ref{sec:proof:`a}.
\end{Proof}

To reduce the upper in \eqref{eq:LB} to the EP bound in \cite{chan19lamination}, consider as in \cite{chan19lamination} the case $A=V$, $H(\RX_{E'})\neq 0$ and $H(\RX_{E`/E'})=0$. Let
\begin{align*}
  `l_{\mcP}(B) &=\begin{cases}
    \frac{1}{\abs{\mcP}-1} & V`/B\in \mcP\\
    0 & \text{otherwise}
  \end{cases} 
\end{align*}
for any partition $\mcP$ of $V$ with $\abs{\mcP}>1$. Then,
\begin{align*}
  `a(`l_{\mcP}) &= \min_{e\in E'} \sum_{B\nsupseteq A: `x(e)\subseteq B} `l_{\mcP}(B)\\
  &=  \frac{1}{\abs{\mcP}-1} \min_{e\in E'} \abs{\Set{V`/B\in \mcP|`x(e)\subseteq B}}\\
  &=  \frac{1}{\abs{\mcP}-1} \min_{e\in E'} `1[\abs{\mcP} - \abs{\Set{C\in \mcP|`x(e)\cap C\neq `0}}`2]\\
  &= 1 - \frac{\max_{e\in E'}\abs{\Set{C\in \mcP|`x(e)\cap C\neq `0}}}{\abs{\mcP}-1}.
\end{align*}
Substituting the above into \eqref{eq:LB} gives the EP bound~\cite[Theorem~4.1]{chan19lamination}.

Next, to reduce to the VP bound in \cite{chan19lamination}, define for $u_V\in `R_+^V$
\begin{align*}
  `l_{u_V}(B) &=\begin{cases}
    \frac{u_i}{u(V)} & i\in V, B\in \Set{\Set{i},V`/\Set{i}}\\
    0 & \text{otherwise}
  \end{cases} 
\end{align*}
Then, 
\begin{align*}
  `a(`l_{u_V}) &= \min_{e\in E'} \sum_{B\nsupseteq A: `x(e)\subseteq B} `l_{u_V}(B)\\
  &\geq  \min_{e\in E'} \sum_{i\in `x(e)|x(e)\subseteq V`/\Set{i}} \frac{u_i}{u(V)}\\
  &=  \min_{e\in E'} \sum_{i\in V`/`x(e)} \frac{u_i}{u(V)}\\
  &=  1-\max_{e\in E'} \frac{u(`x(e))}{u(V)}
\end{align*}
Applying the above to \eqref{eq:LB} 
In particular, with $u_V$ chosen to be the solution to 
\begin{align*}
  `t = \max_{u_V\in `R_+^V:u(`x(e))\leq 1, \forall e\in E} u(V),
\end{align*}
we have
\begin{align*}
  `a(`l_{u_V}) &\geq 1- \max_{e\in E'} \frac{1}{u(V)}\\
  &= 1-\frac1{`t}.
\end{align*}
Applying the above into \eqref{eq:LB}
gives the VP bound~\cite[Theorem~4.3]{chan19lamination}.

\section{Challenges}
\label{sec:cg}

For hypergraphical sources, it is plausible that the upper bound~\eqref{eq:CSrV:ub} of $\CS(r_V)$ in Theorem~\ref{thm:general} may also tight. In this work, we have only shown that the corresponding upper bound~\eqref{eq:CSR:ub2} of $\CS(R)$ in Corollary~\ref{cor:hyp} under total discussion rate constraint instead of individual rate constraints is tight.

If we consider more general sources beyond the hypergraphical sources, however, the bound~\eqref{eq:CSR:ub2} on $\CS(R)$ may be loose. In this section, we give an example of a finite linear source where the bound~\eqref{eq:RSrK:lb} on $\RS$ is loose, and so is \eqref{eq:CSR:ub2}. Nevertheless, it remains plausible that compressed secret key agreement and linear network coding is optimal for general finite linear sources. 

Consider $A=V=\Set{1,2,3,4,5}$ and
\begin{align}
\begin{split}
    \RZ_1 &= \RX_a\\
    \RZ_2 &= \RX_b\\
    \RZ_3 &= \RX_c\\
    \RZ_4 &= (\RX_a,\RX_b,\RX_d)\\
    \RZ_5 &= (\RX_a,\RX_b,\RX_c\oplus \RX_d)
\end{split}\label{eq:ceg}
\end{align}
where $\RX_a,\RX_b,\RX_c,\RX_d$ are uniformly random and independent bits, and $\oplus$ denotes the XOR or binary addition operation.

\begin{Proposition}
    $\RS=3>2\geq `r(\CS)$ for \eqref{eq:ceg} and so the bound~\eqref{eq:RSrK:lb} on the communication complexity is loose.
\end{Proposition}

\begin{Proof}
 It was shown in \cite{chan17oo} that $\CS=1$ and $\RS=\RCO=3$. It remains to show that the lower bound~\eqref{eq:RSrK:lb} is at most $2$, i.e.,
 \begin{align*}
  \inf_{\RW:I(\RW\wedge \RZ_V)-`r(\RW)\geq \CS \kern-2em} \kern-2em \raisebox{2pt}{$I(\RW\wedge \RZ_V)-\CS $}\geq 2
 \end{align*}
 To do so, it suffices to show that a feasible solution $\RW$ to the L.H.S.\ is
 \begin{align*}
  \RW = (\RX_a,\RX_b,\RX_c)  
 \end{align*}
 because then the bound is at most
 \begin{align*}
      I(\RZ_V\wedge \RW) - \CS &= H(\RX_a,\RX_b,\RX_c) - 1 = 2
 \end{align*}
 as desired. 
 
 It remains to show the feasibility, i.e., the following constraint holds,
\begin{align*}
  I(\RW\wedge \RZ_V)-`r(\RW)\geq \CS.
\end{align*}
Note that $I(\RW\wedge \RZ_V)=H(\RX_a,\RX_b,\RX_c)=3$, $\CS=1$, and $`r(\RW)=\min_{r'_V:\eqref{eq:r'_V1}}r'(V)$ by linear programming duality, it suffices to show that 
\begin{align*}
  \min_{r'_V:\eqref{eq:r'_V1}}r'(V) \leq 2.
\end{align*}
In particular, we will argue that a feasible solution with $r'(V)\leq 2$ is
\begin{align*}
  r'_i = \begin{cases}
    1 & i\in \Set{4,5}\\
    0 & i\in \Set{1,2,3}.
  \end{cases}
\end{align*}
More precisely, we will argue that the constraint~\eqref{eq:r'_V1} that
\begin{align*}
  r'(B) \geq I(\RW\wedge \RZ_B|\RZ_{V`/B})\quad \forall B\subsetneq V.
\end{align*}
We can divide all the cases of $B$ as follows:
\begin{itemize}
  \item $4\not\in B$ and $5\not\in B$. Then,
  \begin{align*}
    I(\RW\wedge \RZ_B|\RZ_{V`/B}) \leq H(\RW|\RZ_4,\RZ_5) = 0
  \end{align*}
  and so the constraint holds trivially as $r'_i\geq 0$ for all $i\in V$.
  \item $4\not\in B$ or $5\not\in B$ but not both. Then, 
  \begin{align*}
    I(\RW\wedge \RZ_B|\RZ_{V`/B}) &\leq \max\Set{H(\RW|\RZ_4),H(\RW|\RZ_5)} = 1\\
    r(B)&\geq \min\Set{r_4,r_5}=1
  \end{align*}
  Since and so the constraint holds.
  \item $\Set{4,5}\subseteq B$ but $1\not\in B$ or $2\not \in B$ or $3\not\in B$. Then,
  \begin{align*}
    I(\RW\wedge \RZ_B|\RZ_{V`/B}) &\leq \max_{i\in \Set{1,2,3}}H(\RW|\RZ_i) \leq 2\\
    r(B)&\geq r_4+r_5=2
  \end{align*}
  and so the constraint holds.
\end{itemize}
This completes the proof.
\end{Proof}

\appendix


\subsection{Proof of Theorem~\ref{thm:general}}
\label{sec:proof:thm:general}

We first derive \eqref{eq:CSrV:ub2} from \eqref{eq:CSrV:ub1}. By the linear programming duality, we can rewrite \eqref{eq:CSrV:ub1} as
\begin{align}
  &\max_{\RW} \min_{`l,`m_V} I(\RW\wedge \RZ_V) \\
  &\quad - \sum_{B\nsupseteq A} `l(B) I(\RW\wedge \RZ_B|\RZ_{V`/B}) + \sum_{i\in V} `m_i r_i \label{eq:CS:dual:1}
\end{align} 
where $`l:\Set{B\subseteq V\mid B\nsupseteq A}\to `R_+$ and $`m_V\in `R_+^V$ are subject to the constraint
\begin{align*}
  \sum_{B\nsupseteq A:i\in B} `l(B)-`m_i&=1\quad \forall i\in V.
\end{align*}
The above constraint holds if and only if
\begin{align}
  `l &\in \bar{`L} \label{eq:CS:dual:2}\\
  `m_i&=\sum_{B\supseteq A:i\in B} `l(B)-1,\label{eq:CS:dual:3}
\end{align}
which implies
\begin{align}
  \sum_{i\in V} `m_i r_i 
  &= \sum_{i\in V} \sum_{B\nsupseteq A:i\in B} [`l(B)-1] r_i\notag\\
  &= \sum_{B\nsupseteq A} `l(B)r(B) -r(V).\label{eq:CS:dual:4}
\end{align}
Substituting \eqref{eq:CS:dual:4} into \eqref{eq:CS:dual:1} gives \eqref{eq:CSrV:ub2}, and we need only impose \eqref{eq:CS:dual:2} but not \eqref{eq:CS:dual:3} as \eqref{eq:CSrV:ub2} does not depend on $`m_V$.

To prove \eqref{eq:CSrV:ub1}, we first consider the case without randomization. More precisely, let $\CS^{\op{NR}}(r_V)$ be the secrecy capacity~\eqref{eq:CSrV} but with no randomiziation, i.e., with \eqref{eq:U} replaced by $\RU_V=0$. We want to the that $\CS^{\op{NR}}(r_V)$ is bounded by \eqref{eq:CSrV:ub1}.  
For $i\in V$, let
\begin{align*}
  r_i'&:= \frac1n \sum_{1\leq j\leq r:i_j=i} H(\tRF_{j}|\tRF^{j-1}).
\end{align*}
It follows that
\begin{align} 
  r_i' &\leq \frac1n \sum_{1\leq j\leq r:i_j=i} H(\tRF_{j}|(\tRF_{j'}\mid j'\leq j,i_{j'}=i))\notag\\
  &=\frac1n H(\RF_i) \notag\\
  r_i' &\leq r_i + `d_n\label{eq:rr'},
\end{align}
where the first inequality is because conditioning reduces entropy; the last two steps are by \eqref{eq:Fi} and respectively \eqref{eq:ri} for some $`d_n\to 0$ as $n\to 0$. Furthermore, for $B\subseteq V$,
\begin{align}
  r'(B) &\geq \frac1n \sum_{i\in B} \sum_{1\leq j\leq r: i_j=i} H(\tRF_j\mid \tRF^{j-1},\RZ^n_{V`/B})\notag\\
  &= \frac1n \sum_{i\in V} \sum_{1\leq j\leq r: i_j=i} H(\tRF_j\mid \tRF^{j-1},\RZ^n_{V`/B})\notag\\
  &= \frac1n H(\RF\mid \RZ_{V`/B}^n)\label{eq:r'B}
\end{align}
where the first inequality is again because conditioning reduces entropy; the second equality is because the terms in the summation is $0$ for $i\in V`/B$ by the definition~\eqref{eq:tF} of $\tRF_j$.
For $B=V$, the first inequality holds with equality and so
\begin{align}
  r'(V)=\frac1n H(\RF).\label{eq:r'V}
\end{align}

Next, we single-letterize the key rate and discussion rate as follows. Let
\begin{align}
  \RW_j:=(\RK,\RF,\RZ_V^{j-1})\label{eq:Wj}
\end{align}
and $\RJ$ be a random variable uniformly distributed over $\Set{1,\dots,n}$ and independent of all other random variables. By the secrecy constraint~\eqref{eq:secrecy}, 
\begin{align}
  \log \abs{K} 
  &\leq \underbrace{H(\RK\mid \RF)}_{H(\RK,\RF) - H(\RF)} + n`d_n\notag\\[-1em]
  &\utag{a}= \underbrace{I(\RK,\RF\wedge \RZ_V^n)}_{\mathrlap{\begin{aligned}
    &=\sum_{j=1}^n I(\RK,\RF\wedge \RZ_{Vj}|\RZ_V^{j-1})\\
    &\utag{d}=\sum_{j=1}^n I(\underbrace{\RK,\RF,\RZ_V^{j-1}}_{=\RW_j} \wedge \RZ_{Vj})\\
    &\utag{e}=nI(\RW_{\RJ} \wedge \RZ_{V\,\RJ}) 
  \end{aligned}}}  + \overbrace{H(\RK,\RF|\RZ_V^n)}^{\utag{b}\leq n`d_n}- \overbrace{H(\RF)}^{\utag{c}= nr'(V)} + n`d_n\notag\\
  \frac{\log \abs{K}}n & = I(\RW_{\RJ} \wedge \RZ_{V\,\RJ}) - r'(V) + 2`d_n\label{eq:sl:K}
\end{align}
for some $`d_n\to 0$ as $n\to 0$. \uref{a} is because $H(\RK,\RF)=I(\RK,\RF\wedge \RZ_V^n)+H(\RK,\RF|\RZ_V^n)$. \uref{b} is by the recoverability constraint~\eqref{eq:recover} and Fano's inequality while \uref{c} is by \eqref{eq:r'V}. \uref{d} is because $I(\RZ_V^{j-1} \wedge \RZ_{Vj}|\RK,\RF)=0$ by the memorylessness of the random source $\RZ_V$. \uref{e} follows from the definition~\eqref{eq:Wj} of $\RW_j$ and $\RJ$. 

Similarly, by \eqref{eq:recover} and Fano's inequality,
\begin{align}
  \kern-.2em \underbrace{H(\RF|\RZ_{V`/B}^n)}_{\utag{f}\leq nr'(B)} &\geq \overbrace{H(\RK,\RF\mid \RZ_{V`/B}^n)}^{= H(\RK,\RF) - I(\RK,\RF\wedge \RZ_{V`/B}^n)}-n`d_n\notag\\[-2em]
  &\geq  \overbrace{I(\RK,\RF\wedge \RZ_V^n)}^{\utag{g}=nI(\RW_{\RJ} \wedge \RZ_{V\,\RJ})}-\underbrace{I(\RK,\RF\wedge \RZ_{V`/B}^n)}_{\mathclap{\begin{aligned}
    &=\sum_{j=1}^n I(\RK,\RF,\RZ_{V`/B}^{j-1}\wedge \RZ_{V`/B\,j})\\
    &\leq \sum_{j=1}^n I(\underbrace{\RK,\RF,\RZ_V^{j-1}}_{=\RW_j} \wedge \RZ_{V`/B\,j})\\
    &\utag{h}=nI(\RW_{\RJ} \wedge \RZ_{V`/B\,\RJ}) 
  \end{aligned}}}
  -n`d_n.\notag\\
  \kern-.2em r'(B) &\geq I(\RW_{\RJ} \wedge \RZ_{V\,\RJ})-I(\RW_{\RJ} \wedge \RZ_{V`/B\,\RJ})-`d_n,\kern-.3em\label{eq:sl:F}
\end{align}
for some $`d_n\to 0$ as $n\to 0$.
\uref{f} is by \eqref{eq:r'B}. \uref{g} follows from \uref{e} while \uref{h} follows from the same argument for \uref{e}. \eqref{eq:sl:F} follows from \eqref{eq:r'B}.

Since \eqref{eq:rr'}, \eqref{eq:sl:K}, and \eqref{eq:sl:F} holds for any secret agreement scheme $(\RK,\RF)$, we have the desired bound~\eqref{eq:CSrV:ub1} on $\CS^{\op{NR}}(r_V)$ by setting $n\to `8$ and noting that $P_{\RZ_{V\,\RJ}}=P_{\RZ_V}$. 

It remains to extend the bound~\eqref{eq:CSrV:ub1} to the general case with randomization~\eqref{eq:U} where $\RU_V$ not necessarily deterministic. Let 
\begin{align}
  \RZ'_i:=(\RU_i,\RZ_i)\quad \text{for }i\in V.\label{eq:CS:ub:Z'}
\end{align}
We have
\begin{align*}
  \CS[\RZ_V](r_V) \leq \CS^{\op{NR}}[\RZ'_V](r_V)
\end{align*}
because a secret key agreement scheme with randomization for $\RZ_V$ is also a secret key agreement scheme with no randomization but for $\RZ'_V$. It suffices to show that the R.H.S.\ is upper bounded by \eqref{eq:CSrV:ub}.

Applying the bound~\eqref{eq:CSrV:ub2} with the source $\RZ'_V$ instead of $\RZ_V$ for the secrecy capacity with no randomization, we have 
\begin{align*}
  \CS^{\op{NR}}[\RZ'_V](r_V) 
  &\leq \max_{\RW}\min_{`l\in \bar{`L}} I(\RW \wedge \RZ'_V) -r(V) \notag\\
  &\kern1em -\sum_{B\nsupseteq A} `l(B)`1[I(\RW\wedge \RZ'_B|\RZ'_{V`/B}) - r(B)`2].
\end{align*}
It suffices to show that the above bound is upper bounded by \eqref{eq:CSrV:ub2}, i.e., the bound above remains valid after replacing $\RZ'_V$ by $\RZ_V$. In particular, we will show that
\begin{align}
  \begin{split}
  &\kern-2em I(\RW\wedge \RZ'_V) - \sum_{B\nsupseteq A} `l(B) I(\RW\wedge \RZ'_B|\RZ'_{V`/B})\\
  &\leq I(\RW\wedge \RZ_V) - \sum_{B\nsupseteq A} `l(B) I(\RW\wedge \RZ_B|\RZ_{V`/B}).
  \end{split}\label{eq:CSCS'}
\end{align}

Consider $V=\Set{1,\dots,m}$ without loss of generality and define
\begin{align*}
  \tdr_i:=I(\RW\wedge \RU_i|\RU^{i-1},\RZ_{V}).
\end{align*}
It follows that
\begin{align}
  \tdr(B)&=\sum_{i\in B} \underbrace{I(\RW\wedge \RU_i|\RU^{i-1}\RZ_{V})}_{\mathclap{\kern4em=H(\RU_i|\RU^{i-1},\RZ_{V})-H(\RU_i|\RU^{i-1},\RZ_{V},\RW)}}\notag\\
  &\leq  \sum_{i\in B} \overbrace{I(\RW\wedge \RU_i|\RU^{i-1},\RU_{V`/B},\RZ_{V})}^{\mathclap{\kern2em=H(\RU_i|\RU^{i-1},\RU_{V`/B},\RZ_{V})-H(\RU_i|\RU^{i-1},\RU_{V`/B},\RZ_{V},\RW)}}\notag\\
  &= I(\RW\wedge \RU_B|\RU_{V`/B},\RZ_V)\label{eq:tdrB}
\end{align}
where the inequality holds with equality if $B=V$. This is because $H(\RU_i|\RU^{i-1},\RZ_{V})=H(\RU_i|\RU^{i-1},\RU_{V`/B},\RZ_{V})$ by the independence assumption~\eqref{eq:U}, and $H(\RU_i|\RU^{i-1},\RZ_{V},\RW)\geq H(\RU_i|\RU^{i-1},\RU_{V`/B},\RZ_{V},\RW)$ with equality if $B=V$. It follows that
\begin{align}
  &\kern-1em I(\RW\wedge \RZ'_B|\RZ'_{V`/B}) \notag\\
  &= I(\RW\wedge \RU_B,\RZ_B|\RU_{V`/B},\RZ_{V`/B})\notag\\[-1em]
&= \underbrace{I(\RW\wedge \RZ_B|\RU_{V`/B},\RZ_{V`/B})}_{\mathclap{\kern10em \begin{aligned}
  &= H(\RZ_B|\RU_{V`/B},\RZ_{V`/B}) - H(\RZ_B|\RU_{V`/B},\RZ_{V`/B},\RW)\notag\\
  &\geq H(\RZ_B|\RZ_{V`/B}) - H(\RZ_B|\RZ_{V`/B},\RW) \quad\text{by \eqref{eq:U}}\notag\\
  &= I(\RW\wedge \RZ_B|\RZ_{V`/B})
\end{aligned}}}
  + \overbrace{I(\RW\wedge \RU_B|\RU_{V`/B},\RZ_V)}^{\geq \tdr(B)\quad\text{by \eqref{eq:tdrB}}}\notag\\
  &\geq I(\RW\wedge \RZ_B|\RZ_{V`/B}) + \tdr(B).\label{eq:IWZ'}
\end{align}
Again, the inequalities above holds with equality if $B=V$. 

Applying \eqref{eq:IWZ'} to the L.H.S.\ of \eqref{eq:CSCS'} and subtract the resulting lower bound from the R.H.S.\ of \eqref{eq:CSCS'}, we have
\begin{align*}
  &\kern-2em \text{L.H.S.\ of \eqref{eq:CSCS'}}-\text{R.H.S.\ of \eqref{eq:CSCS'}}\\
  &\geq \sum_{B\nsupseteq A} `l(B) \sum_{i\in B} \tdr(B) - \tdr(V)\\
  &= \sum_{i\in V} \underbrace{\sum_{B\nsupseteq A:i\in B}}_{\geq 1\quad\text{by \eqref{eq:fc}}} `l(B)  \tdr(B) -\tdr(V)\\
  &\geq \tdr(V)-\tdr(V) = 0,
\end{align*}
which implies \eqref{eq:CSCS'} as desired and therefore completes the proof.

\subsection{Proof of Corollary~\ref{cor:general}}
\label{sec:proof:cor:general}

We first derive the bound~\eqref{eq:CS:ub} on the unconstrained capacity by \eqref{eq:CSR:ub2}. By the definition~\ref{eq:I_l} of $I_{`l}$,
\begin{align}
  &\kern-2em I_{`l}(\RZ_V) - I_{`l}(\RZ_V|\RW)\notag\\
  &= H(\RZ_V) - \sum_{B\nsupseteq A} H(\RZ_B|\RZ_{V`/B})\notag\\
  &\kern3em -H(\RZ_V|\RW) + \sum_{B\nsupseteq A} H(\RZ_B|\RZ_{V`/B},\RW)\notag\\
  &= I(\RW \wedge \RZ_V) - \sum_{B\nsupseteq A} `l(B) I(\RW\wedge \RZ_B|\RZ_{V`/B}).\label{eq:I_l:exp}
\end{align}
Hence, by~\eqref{eq:CSR:ub2}, for $R\geq 0$,
\begin{align*}
  \CS(R) &\leq \sup_{\RW} I(\RW \wedge \RZ_V) - `r(\RW)\\
  &= \sup_{\RW}\min_{`l} H(\RZ_V) - \sum_{B\nsupseteq A} H(\RZ_B|\RZ_{V`/B})\\
  &\leq \sup_{\RW}\min_{`l} I_{`l}(\RZ_V) - \underbrace{I_{`l}(\RZ_V|\RW)}_{\geq 0\quad \text{by Shearer Lemma~(see \eqref{eq:shearer})}}\\
  &\leq \min_{`l} I_{`l}(\RZ_V)
\end{align*}
which gives \eqref{eq:CS:ub} as desired since the bound does not depend on $R$.

Next, we derive the upper bounds~\eqref{eq:CSR:ub1} and \eqref{eq:CSR:ub2} on $\CS(R)$ from the upper bound~\eqref{eq:CSrV:ub} on $\CS(r_V)$ as follows: For $R\geq 0$. By \eqref{eq:CSR},
\begin{align*}
  \CS(R) &=\sup\Set{\CS(r_V)|r(V)\leq R}\\
  &\utag{a}\leq \sup\Set{I(\RW\wedge \RZ_V)-r'(V)|r(V)\leq R,\text{\eqref{eq:r'_V}}}\\
  &\utag{b}\leq \sup\Set{I(\RW\wedge \RZ_V)-r'(V)|r'(V)\leq R,\text{\eqref{eq:r'_V2}}}\\
  &\utag{c}= \sup\Set{I(\RW\wedge \RZ_V)-\min_{r'_V:\text{\eqref{eq:r'_V2}}} r'(V)|\min_{r'_V:\text{\eqref{eq:r'_V2}}} r'(V)\leq R}\\
  &\utag{d}= \sup\Set{I(\RW\wedge \RZ_V)-`r(\RW)|`r(\RW)\leq R},
\end{align*}
which gives \eqref{eq:CSR:ub2} as desired. \uref{b} also gives \eqref{eq:CSR:ub1}.
\uref{a} is by \eqref{eq:CSrV:ub}. \uref{b} is because \eqref{eq:r'_V2} and $r(V)\leq R$ imply $r'(V)\leq R$. \uref{c} is because it is optimal to choose $r'_V$ to minimize $r'(V)$. \uref{d} is by the linear programming duality and the definition~\eqref{eq:rho} of $`r(\RW)$.

Finally, we can derive the lower bound~\eqref{eq:RSrK:lb} on $\RS(r_{\opK})$ from \eqref{eq:CSR:ub2} as follows: By \eqref{eq:RSrK},
\begin{align*}
  \RS(r_{\opK}) 
  &= \inf\Set{R\geq 0|\CS(R)\geq r_{\opK}}\\
  &\utag{e}= \inf\Set{R\geq 0|I(\RW\wedge \RZ_V) - `r(\RW) \geq r_{\opK},`r(\RW)\leq `R}\\
  &\utag{f}= \inf\Set{`r(\RW)|I(\RW\wedge \RZ_V) - `r(\RW) \geq r_{\opK}}\\
  &\utag{g}= \inf\Set{`r(\RW)|I(\RW'\wedge \RZ_V) - `r(\RW') = r_{\opK}}\\
  &= \inf\Set{I(\RW'\wedge \RZ_V)-r_{\opK}|I(\RW'\wedge \RZ_V) - `r(\RW') = r_{\opK}}
\end{align*}
which implies \eqref{eq:RSrK:lb} as desired.
\uref{e} is by \eqref{eq:CSR:ub2}. \uref{f} is obtained by setting $`R=`r(\RW)$ without loss of optimality. \uref{g} is because, for any feasible $\RW$ to \uref{f}, we also have a feasible $\RW'$ to \uref{g} and vice versa. E.g., given $\RW$, one can choose
\begin{align*}
  \RW' = \begin{cases}
    (1,\RW) & \RY=1\\
    0 & \RY=0
  \end{cases}
\end{align*}
where $\RY$ is an indicator random variable independent of $(\RZ_V,\RW)$ with $I(\RW'\wedge \RZ_V)=`e$. Note that 
\begin{align*}
  I(\RW'\wedge \RZ_V) &=`eI(\RW'\wedge \RZ_V)\\
  `r(\RW')  &= `e `r(\RW)  
\end{align*}
and so the condition in \uref{g} can be satisfied with some $`e\in [0,1]$ as desired. This completes the proof of Corollary~\ref{cor:general}.

\subsection{Proof of Theorem~\ref{thm:hyp}}
\label{sec:proof:thm:hyp}

In this section, we show that the upper bound~\eqref{eq:CSR:ub2} on the secrecy capacity is tight for hypergraphical sources. Indeed, we show that the lower bound~\eqref{eq:RSrK:lb} on communication complexity is tight by showing that the following sufficient condition for tightness holds for hypergraphical sources.

\begin{Lemma}
  \label{lem:tight}
  The lower bound~\eqref{eq:RSrK:lb} on communication is tight for all $r_{\opK}\in [0,\CS]$ if there exists an optimal solution $\RW$ to \eqref{eq:RSrK:lb} in the form 
  \begin{subequations}
    \label{eq:tight}   
    \begin{align}
      \RW&=(\RQ,\RZ'_V) && \text{such that}\label{eq:tight:W}\\
      I(\RQ \wedge \RZ_V) &= 0 \label{eq:tight:Q}\\
      H(\RZ_i'|\RZ_i,\RQ) &= 0 && \forall i\in V \label{eq:tight:Z'}\\
      H(\RZ_B'|\RZ_{V`/B}',\RQ) &= H(\RZ_B'|\RZ_{V`/B},\RQ) && \forall B\nsupseteq A. \label{eq:tight:ZZ'}
    \end{align}
  \end{subequations}
  Furthermore, compressed secret key agreement~\cite{chan17cska} is optimal in achieving $\RS(r_{\opK})$ and therefore $\CS(R)$~\eqref{eq:CSR:ub2} for $R\geq 0$.
\end{Lemma}

\begin{Proof}
  Consider any optimal solution $\RW$ to \eqref{eq:RSrK:lb} satisfying the condition~\eqref{eq:tight}. We will show that
  \begin{align}
    I(\RW\wedge \RZ_{V`/B}) &= H(\RZ'_{V`/B}|\RQ) \quad \forall B\subseteq V.
    \label{eq:WZ'}
  \end{align}
  Then, the lower bound~\eqref{eq:RSrK:lb} can be written as
  \begin{align*}
    I(\RW\wedge \RZ_V) - `r_{\opK}
    &\utag{a}\geq `r(\RW)\\
    &\utag{b}=\max_{`l\in `L} \sum_{B\nsupseteq A} `l(B) I(\RW\wedge \RZ_B|\RZ_{V`/B})\\
    &\utag{c}= H(\RZ'_V|\RQ) - \min_{`l\in `L} I_{`l}(\RZ'_V|\RQ)\\
    &\utag{d}\geq \RS(r_{\opK})
  \end{align*}
  which is the desired reverse inequality of the bound~\eqref{eq:RSrK:lb}. \uref{a} follows from the constraint on $\RW$ in~\eqref{eq:RSrK:lb}. \uref{b} is by the definition~\eqref{eq:rho} of $`r(\RW)$. \uref{c} is obtained by rewriting
  \begin{align*}
    I(\RW\wedge \RZ_B|\RZ_{V`/B}) &= I(\RW\wedge \RZ_V) - I(\RW\wedge \RZ_{V`/B})\\
    &= H(\RZ'_{V}|\RQ) - H(\RZ'_{V`/B}|\RQ)\\
    &= H(\RZ'_B|\RZ'_{V`/B},\RQ)\\
    \sum_{B\supseteq A} I(\RW\wedge \RZ_B|\RZ_{V`/B}) &= H(\RZ'_V|\RQ) - I_{`l}(\RZ'_V|\RQ)
  \end{align*}
  where the second equality is by \eqref{eq:WZ'} and the last equality is by the definition~\eqref{eq:I_l} of $I_{`l}$. \uref{d} is because the R.H.S.\ of \uref{c} is the discussion rate achievable by the compressed secret key agreement scheme in \cite[Theorem~3]{chan17cska}\footnote{The extension to the case with helpers is straightforward as in \cite{csiszar04,csiszar08}.} to attain a key rate of 
  \begin{align*}
    \min_{`l\in `L} I_{`l}(\RZ'_V|\RQ) 
    &\geq H(\RZ'_V|\RQ) - I(\RW\wedge \RZ_V) + `r_{\opK}= `r_{\opK}.
  \end{align*}
  where the first inequality is by \uref{c} and the second equality is by \eqref{eq:WZ'}.

  It remains to show \eqref{eq:WZ'} as follows. For $B\subseteq V$,
  \begin{align*}
    I(\RW \wedge \RZ_{V`/B})
    &= I(\RZ'_V,\RQ \wedge \RZ_{V`/B})\quad \text{by \eqref{eq:tight:W}}\\
    &= I(\RZ'_V \wedge \RZ_{V`/B}|\RQ)+\underbrace{I(\RQ \wedge \RZ_{V`/B})}_{=0 \quad \text{by \eqref{eq:tight:Q}}}\\
    &= \underbrace{I(\RZ'_{V`/B} \wedge \RZ_{V`/B}|\RQ)}_{=H(\RZ_{V`/B}|\RQ)} + \underbrace{I(\RZ'_B \wedge \RZ_{V`/B}|\RQ,\RZ'_{V`/B})}_{=0} 
  \end{align*}
  as desired where the last equality is because
  \begin{align*}
    &\kern-2em I(\RZ'_B \wedge \RZ_{V`/B}|\RQ,\RZ'_{V`/B}) \\
    &= \underbrace{H(\RZ'_B |\RQ,\RZ'_{V`/B})}_{=H(\RZ'_B |\RQ,\RZ_{V`/B})\quad\text{by \eqref{eq:tight:ZZ'}}} - \underbrace{H(\RZ'_B|\RQ,\RZ'_{V`/B},\RZ_{V`/B})}_{=H(\RZ'_B|\RQ,\RZ_{V`/B})\quad\text{by \eqref{eq:tight:Z'}}}
  \end{align*}
  This completes the proof.
\end{Proof}

Note that $\RW$ defined in \eqref{eq:hyp:sc} satisfies~\eqref{eq:tight} with $\RZ'_i$ defined as in \eqref{eq:hyp:Z'}.
In particular,
\begin{align}
  H(\RZ'_B|\RZ_{V`/B},\RQ) = 
  H(\RZ'_B|\RZ'_{V`/B},\RQ) =
  H(\RX'_{E(B)}|\RQ)\label{eq:Z'X'}
\end{align}
where $E(B)$ is defined in \eqref{eq:EB} as the set of edges that are incident only on nodes within $B$.
By the above lemma, it suffices to show that $\RW$ defined in \eqref{eq:hyp:sc} is an optimal solution to the lower bound~\eqref{eq:RSrK:lb}. 

For simplicity, we rewrite the lower bound~\eqref{eq:RSrK:lb} as $H(\RZ_V) - r_{\opK} - T$ where 
\begin{align}
  T &:= \max\Set{H(\RZ_V|\RW)| I(\RW\wedge \RZ_V) - `r(\RW)\geq r_{\opK}}.\label{eq:T}
\end{align}
Since the above maximization share the same set of solutions $\RW$ as that of the lower bound~\eqref{eq:RSrK:lb}, it suffices to show that $\RW$ defined in \eqref{eq:hyp:sc} is optimal to the above maximization. 

We further rewrite $T$ without changing the set of optimal $\RW$ below:
\begin{align*}
  T &\utag{a}= \max\Set{H(\RZ_V|\RW)|I_{`l}(\RZ_V) - I_{`l}(\RZ_V|\RW)\geq r_{\opK}\,\forall `l\in `L}\\
  &\utag{b}= \max\{H(\RZ_V|\RW)|\\
  &\kern3em{\textstyle\sum_{B\nsupseteq A}} `l(B) H(\RZ_{V`/B}|\RW)\leq I_{`l}(\RZ_V) -r_{\opK} \\
  &`1.\kern3em +{\textstyle`1(\sum_{B\nsupseteq A} `l(B)-1`2)} H(\RZ_V|\RW)
  \,\forall `l\in `L`2\}\\
  &\utag{c}= \max\{H(\RZ_V|\RW)|\\
  &\kern3em{\textstyle\sum_{B\nsupseteq A}} `l(B) H(\RZ_{V`/B}|\RW)\leq `a(`l)
  \,\forall `l\in `L\}\\
  &\utag{d}= \max\{H(\RX_E|\RW) | \overbrace{0\leq H(\RX_e|\RW)\leq H(\RX_e)\,\forall e\in E}^{\text{trivial}},\\
  &\kern3em \sum_{B\nsupseteq A}`l(B)H(\RX_{E(V`/B)}|\RW)\leq `a(`l)\,\forall `l\in `L\}
\end{align*}
where $E(\cdot)$ is defined in \eqref{eq:EB}, and
\begin{align}
  `a(`l)&:=I_{`l}(\RZ_V) -r_{\opK} +{\textstyle`1(\sum_{B\nsupseteq A} `l(B)-1`2)} T.\label{eq:`a}
\end{align}  
\begin{itemize}
  \item \uref{a} is obtained by rewriting the constraint in \eqref{eq:T} as
  \begin{align*}
   r_{\opK}&\leq  I_{`l}(\RZ_V) - I_{`l}(\RZ_V|\RW) && \forall `l\in `L,
  \end{align*}
  because 
  \begin{align*}
    &\kern-1em I(\RW\wedge \RZ_V) - `r(\RW) \\
    &= \min_{`l\in `L} \underbrace{I(\RW\wedge \RZ_V) - \sum_{B\nsupseteq A} `l(B) I(\RW\wedge \RZ_B|\RZ_{V`/B})}_{=I_{`l}(\RZ_V) - I_{`l}(\RZ_V|\RW)\text{ by \eqref{eq:I_l:exp}}}.
  \end{align*}
  \item \uref{b} is obtained by expanding $I_{`l}(\RZ_V|\RW)$ by its definition~\eqref{eq:I_l}.
  \item $\leq$ for \uref{c} is because any optimal solution $\RW$ to \uref{b} is feasible to \uref{c}.
  To explain $\geq$ for \uref{c}, suppose to the contrary that there exists a solution $\RW$ to \uref{c} with $H(\RZ_V|\RW)>T$. Then, that solution is also feasible to \uref{b} because the constraint of \uref{c} with $T<H(\RZ_V|\RW)$ implies the constraint in \uref{b}, which is the desired contradiction since the maximum in~\uref{b} equals $T$.
  \item \uref{d} is because $\RZ_{V`/B}=\RX_{E(V`/B)}$. Note that the additional constraints in \uref{d} but not \uref{c} hold trivially as the edge variables have bounded supports. 
\end{itemize}

It remains to argue that $\RW$ defined in \eqref{eq:hyp:sc} is an optimal solution to \uref{d}. More precisely, we show that this choice of $\RW$ achieves the maximum given by the linear program
\begin{subequations}
  \label{eq:T'}
  \begin{align}
    T'&:= \max\{y(E)\mid 0\leq y_e \leq H(\RX_e)\,\forall e\in E\label{eq:T'1}\\
    &\kern3em \sum_{B\nsupseteq A} `l(B) y(E(V`/B))\leq `a(`l)\,\forall `l\in `L\}\label{eq:T'2}
  \end{align}
\end{subequations}
where $y(E'):=\sum_{e\in E'} y_e$ for $E'\subseteq E$ as usual. 
\begin{itemize}
  \item To show $T\leq T'$, consider the dual of the linear program in \uref{e}, i.e.,
  \begin{align}
    \kern-1em T'&= \min_{`m:E\cap `L\to `R_+} \sum_{e\in E} `m(e) H(\RX_e) + \sum_{`l\in `L} `m(`l)`a(`l) \,\text{such that}\kern-2.5em\notag\\
    &`m(e)+\sum_{`l\in `L} `m(`l)\sum_{B\nsupseteq A: e\in E(B)} `l(B)\geq 1\quad \forall e\in E.\label{eq:T':dual}
  \end{align}
  For any optimal solution $`m$, since $H(\RX_{E'}|\RW)$ is submodular in $E'\subseteq E$, i.e.,
  \begin{align*}
    &\kern-1em H(\RX_{E_1}|\RW)+H(\RX_{E_2}|\RW)\\
    &\geq H(\RX_{E_1\cap E_2}|\RW)+H(\RX_{E_1\cup E_2}|\RW)\quad \forall E_1,E_2\subseteq E,
  \end{align*}
  we have by Edmonds' greedy algorithm (see \cite[Lemma~A.1]{chan19lamination}) that, for any solution $\RW$ to \uref{d},
  \begin{align*}
    T=H(\RX_E|\RW) &\leq \sum_{e\in E} `m(e) \underbrace{H(\RX_e|\RW)}_{\leq H(\RX_e)} \\
    &\kern1em + \sum_{`l\in `L} `m(`l) \underbrace{\sum_{B\nsupseteq A} `l(B) H(\RX_{E(B)}|\RW)}_{\leq `a(`l)\quad \text{by \eqref{eq:T'}}}\\
    &\leq T'\quad \text{as desired by \eqref{eq:T'2}.}
  \end{align*}

  \item To show $T\geq T'$, consider any optimal solution $y_E$ to~\eqref{eq:T'} and define  
  $\RW$ by \eqref{eq:hyp:sc} with
  \begin{align}
    x_e := H(\RX_e) - y_e \quad \forall e\in E.\label{eq:y}
  \end{align}
  The above definition of $\RW$ is valid because \eqref{eq:T'1} implies $x_e/H(\RX_e)\in [0,1]$, which is a valid probability~$P_{\RQ_e}$ in \eqref{eq:hyp:Q}. For $E'\subseteq E$,
  \begin{align}
    \label{eq:yE'}
    \begin{split}
    H(\RX_{E'}|\RW) &= H(\RX_{E'}|\RQ_E,\RX'_E)\\
    &= \sum_{e\in E'} H(\RX_e|\RQ_e,\RX'_e)\\
    &= \sum_{e\in E'} \underbrace{H(\RX_e|\RQ_e)}_{=H(\RX_e)}-\underbrace{H(\RX'_e|\RQ_e)}_{=x_e}\\
    &= y(E').
    \end{split}
\end{align}
  The first equality is by \eqref{eq:hyp:W} that $\RW=(\RQ_E,\RX'_E)$. The second equality is by the independence of $(\RQ_e,\RX'_e,\RX_e)$ for $e\in E$, which follows from \eqref{eq:hyp:X'} and \eqref{eq:hyp:Q}. The third equality is by the fact that $H(\RX'_e|\RX_e,\RQ_e)=0$ by \eqref{eq:hyp:X'}. The last equality is by the definition~\eqref{eq:y} of $y_e$.

  It follows that the constraint in \uref{d} holds, i.e., for all $`l\in `L$,
  \begin{align*}
    \sum_{B\nsupseteq A}`l(B) \underbrace{H(\RX_{E(V`/B)}|\RW)}_{=y(E(V`/B))\quad \text{by \eqref{eq:yE'}}}
    &\leq `a(`l)
  \end{align*}
  by \eqref{eq:T'2}. By the feasibility of $\RW$ to \uref{d},
  \begin{align*}
    T \geq H(\RX_E|\RW) = y(E) = T'
  \end{align*}
  as desired. The first equality is by \eqref{eq:yE'} and the last is by the optimality of $y_E$ to \eqref{eq:T'}.
\end{itemize}
This completes the proof.

\subsection{Proof of Corollary~\ref{cor:hyp}}
\label{sec:proof:cor:hyp}

With the definition~\eqref{eq:hyp} of $\RW$ and the definition~\eqref{eq:hyp:Z'} of $\RZ'_V$, we have
\begin{align}
  \begin{split}
  I(\RW\wedge \RZ_B|\RZ_{V`/B}) &= I(\RW\wedge \RZ_V) - I(\RW\wedge \RZ_{V`/B})\\
  &\utag{a}= H(\RZ'_V|\RQ) - H(\RZ'_{V`/B}|\RQ)\\
  &= H(\RZ'_B|\RZ'_{V`/B},\RQ)\\
  &\utag{b}= H(\RX'_{E(B)}|\RQ) \\
  &\utag{c}= \sum_{e\in E(B)} H(\RX'_e|\RQ) \\
  &\utag{d}= x(E(B)) 
  \end{split}\label{eq:xI}
\end{align}

where \uref{a} is by \eqref{eq:WZ'}, \uref{b} is by \eqref{eq:Z'X'}, \uref{c} is by \eqref{eq:hyp:X'}, and \uref{d} is by \eqref{eq:hyp:Q}. 

Applying the above equality to the capacity upper bound~\eqref{eq:CSR:ub1} and \eqref{eq:CSR:ub2} gives \eqref{eq:hyp:CSR1} and \eqref{eq:hyp:CSR2} respectively. Similarly, \eqref{eq:hyp:RS} follows from \eqref{eq:RSrK:lb}.

By the ellipsoid method~\cite{schrijver02}, to show that the linear programs are polynomial-time solvable, it suffices to show that the following separation oracle for \eqref{eq:hyp:CSR1} is. 
\begin{align*}
  0\leq \max\{
    &R-r(V),\\
    &\min_{e\in E} x_e,\\
    &\min_{e\in E} H(\RX_e) - x_e,\\
    &\min_{i\in A} \min_{B\subseteq V`/\Set{i}} r(B) - x(E(B))\}.
\end{align*}
In particular, in the last expression, $\min_{i\in A} \min_{B\subseteq V`/\Set{i}}$ is equivalent to $\min_{B\subseteq V:B\nsupseteq A}$. 
It is straightforward to verify that the above inequality holds if and only if $(r_V,x_E)$ is a feasible solution to \eqref{eq:hyp:CSR}. It suffices to show that the last minimization 
\begin{align}
  \min_{B\subseteq V`/\Set{i}} r(B) - x(E(B)) \label{eq:SFM}
\end{align}
is polynomial-time solvable despite having exponentially many constraints. To argue this, note that by \eqref{eq:xI}
\begin{align}
  r(B)-x(E(B)) &= \sum_{i\in B} r_i -  H(\RZ'_V|\RQ) - H(\RZ'_{V`/B}|\RQ),\label{eq:submodular}
\end{align}
which is submodular in $B$ by the submodularity of entropy~\cite{fujishige78}.
Hence, \eqref{eq:SFM} is a submodular function minimization over a lattice family, namely the boolean lattice, which is known to be strongly polynomial-time solvable~\cite{fujishige05}. This completes the proof.

\subsection{Proof of Proposition~\ref{pro:HS}}
\label{sec:proof:HS}

Consider $S\subseteq V:\abs{A`/S}\geq 2$ and fractional partition $`l':\Set{B\subseteq V`/S: B\nsupseteq A`/S}\to `R_+$ of $V`/S$ as stated in the proposition. To prove the necessary condition~\eqref{eq:HS} for $\CS(r_V)\geq r_{\opK}$ and $r_V\in `R_+^V$, it suffices to show the following upper bound
\begin{subequations}
  \label{eq:CSrV:ub3}
  \begin{align}
    \CS(r_V) &\leq I_{`l'}(\RZ_{V`/S}) + (`D-1) r(S)\quad\text{where}\\
    `D&:=\sum_{B\nsupseteq A`/S} `l'(B).\label{eq:`D}
  \end{align}      
\end{subequations}
In particular, we will show that the above bound is given by the upper bound~\eqref{eq:CSrV:ub2} with $`l\in \bar{`L}$ defined in terms of $`l'$ as follows:
\begin{align}
  `l(B) = \begin{cases}
    `l'(B`/S) & B\subseteq V: B\supseteq S\\
    0 & \text{otherwise.}
  \end{cases}\label{eq:HS:`l}
\end{align}
$`l\in bar{`L}$, i.e., \eqref{eq:fc} holds, because
\begin{align}
  \sum_{B\nsupseteq A:i\in B} `l(B) -1 &=
  \begin{cases}
    \sum_{B\nsupseteq A`/S:i\in B} `l'(B)-1 & i\in V`/S\\
    \sum_{B\nsupseteq A`/S} `l'(B) - 1 & \text{otherwise.}
  \end{cases}\notag\kern-3em\\
  &=     
  \begin{cases}
    0 & i\in V`/S, \quad \text{by \eqref{eq:HS:`l'}}\\
    `D - 1 & \text{otherwise,}\quad\text{by \eqref{eq:`D}}
  \end{cases} \label{eq:HS:sum`l} \\
  &\geq 0.\notag
\end{align}
The last inequality is because, for any $i\in V`/S$, we have $`D\geq\sum_{B\nsupseteq A`/S:i\in B} `l(B) =1$. 

Note that the upper bound~\eqref{eq:CSrV:ub2} remains a valid upper bound for any given choice of $`l$, i.e.,
\begin{align*}
  \CS(r_V) &\leq \overbrace{I(\RW \wedge \RZ_V)}^{`(1)} - \overbrace{\sum_{B\nsupseteq A} `l(B) I(\RW\wedge \RZ_B|\RZ_{V`/B})}^{`(2)}\\
  &\kern1em + \underbrace{`1[\sum_{B\nsupseteq A} `l(B) r(B) -r(V)`2]}_{`(3)}.
\end{align*}
With the choice of $`l$ given by \eqref{eq:HS:`l},
\begin{align*}
  `(3) &=  \underbrace{\sum_{B\nsupseteq A}`l(B) \sum_{i\in B} `l(B) r_i}_{\sum_{i\in V} \sum_{B\nsupseteq A: i\in B} `l(B) r_i} - \sum_{i\in V} r_i\\
    &= \sum_{i\in V} r_i `1[\sum_{B\nsupseteq A: i\in B} `l(B) - 1`2]\\
    &= (`D-1) r(S).
\end{align*}
where the last equality is by \eqref{eq:HS:sum`l}.
\begin{align*}
`(2) &= \sum_{B\nsupseteq A} `l(B) I(\RW\wedge \RZ_B|\RZ_{V`/B})\\
&= \sum_{B\nsupseteq A`/S} `l'(B) \underbrace{I(\RW\wedge \RZ_{B\cup S}|\RZ_{V`/(B\cup S)})}_{\mathrlap{=I(\RW\wedge \RZ_B|\RZ_{V`/S`/B})+I(\RW\wedge \RZ_S|\RZ_{V`/S})}}\\
&= `D I(\RW\wedge \RZ_S|\RZ_{V`/S})+ \sum_{B\nsupseteq A`/S} `l'(B) I(\RW\wedge \RZ_B|\RZ_{V`/S`/B}),
\end{align*}
where the last equality is again by \eqref{eq:HS:sum`l}. 
\begin{align*}
`(1)-`(2) &= \underbrace{I(\RW\wedge \RZ_V)}_{\mathrlap{
  =I(\RW\wedge \RZ_{V`/S})+I(\RW\wedge \RZ_S|\RZ_{V`/S})}} 
  + `(2)\\
&= \overbrace{(1-`D-1) I(\RW\wedge \RZ_S|\RZ_{V`/S})}^{\geq 0\quad \because `D\geq 1}\\
&\kern1em + \underbrace{I(\RW\wedge \RZ_{V`/S}) - \sum_{B\nsupseteq A`/S} `l'(B) I(\RW\wedge \RZ_B|\RZ_{V`/S`/B})}_{=I_{`l'}(\RZ_{V`/S})-I_{`l'}(\RZ_{V`/S}|\RW)\quad \text{by \eqref{eq:I_l:exp}}}\\
&\leq I_{`l'}(\RZ_{V`/S})-\underbrace{I_{`l'}(\RZ_{V`/S}|\RW)}_{\mathrlap{\geq 0\quad \text{by Shearer Lemma (see \eqref{eq:shearer}).}}}
\end{align*}
Hence, $`(1)-`(2)+`(3)$ can be upper bounded by \eqref{eq:CSrV:ub3}, which completes the proof.

\subsection{Proof of Proposition~\ref{pro:LB}}
\label{sec:proof:LB}

Consider the characterization~\eqref{eq:hyp:CSR2} of $\CS(R)$. It is optimal to set
\begin{align*}
  x_e &= H(\RX_e) \quad \forall e\in E`/E'
\end{align*}
because $`r$ does not depend on the above $x_e$'s as $`x(e)\supseteq A$ for $e\in E`/E'$. \eqref{eq:hyp:CSR2} can be rewritten as
\begin{subequations}
  \label{eq:LB1}
  \begin{align}
    \CS(R) - H(\RX_{E`/E'}) &= \max\{ x(E') - `r | `r \leq R \label{eq:LB1:R}\\
    &\kern1em`r = \max_{`l\in `L} \sum_{B\nsupseteq A} `l(B) x(E(B))\label{eq:LB1:`r}\\
    &\kern1em 0\leq x_e\leq H(\RX_e) \quad \forall e\in E'\label{eq:LB1:x}
  \}.
  \end{align}
\end{subequations}
The above maximization is at least $0$, for instance, by choosing $x_e=0$ for all $e\in E'$. This implies the first inequality in \eqref{eq:LB} as desired. It remains to show the second inequality.

Consider the case $H(\RX_{E'})= 0$ and so $`a=1$ by definition~\eqref{eq:`a}. Then, in \eqref{eq:hyp:CSR2}, we must have $x(E')=`r=0$ because $x(E(B))\leq x(E') \leq H(\RX_{E'})=0$ for all $B\subseteq V$. Hence, the maximum in \eqref{eq:LB1} is $0$, which implies \eqref{eq:LB} as desired since $`a=1$ by the definition~\eqref{eq:`a}. 

Consider the remaining case $H(\RX_{E'})\neq 0$ and any optimal solution $x_{E'}$ and $`r$ to \eqref{eq:LB1} such that $`r>0$. To show that the choice of optimal solution is possible, note that any feasible solution satisfies
\begin{align}
  x(E')\geq \sum_{B\nsupseteq A} `l(B) x(E(B)) \quad \forall x_{E'}\in `R_+^{E'},`l\in `L \label{eq:shearer1}
\end{align}
by the Shearer Lemma. (See \eqref{eq:shearer} and \eqref{eq:submodular}.) Suppose to the contrary that any optimal solution must have $`r=0$. By \eqref{eq:LB1:`r}, we must have $x_e=0$ for all $e\in E'$ which implies $x(E')-`r=0$. This is a contradiction because it is possible to choose $x_e>0$ for some $e\in E'$ as $H(\RX_{E'})\neq 0$, and such choice is also optimal as $x(E')-`r\geq 0$ by \eqref{eq:shearer1}.

Next, with the optimal solution $x_{E'}$ and $`r>0$, \eqref{eq:LB1} becomes
\begin{align*}
  \CS(R) - H(\RX_{E`/E'}) &\leq x(E') - `r\\
  &= `1[\frac1{`r/x(E')} - 1`2] `r\\
  &\utag{a}\leq `1[\frac1{`r/x(E')} - 1`2] R
\end{align*}
with equality if $`r=R$. It suffices to show $\frac{`r}{x(E')}\geq `a(`l)$, which gives the second inequality in \eqref{eq:LB1}.
By \eqref{eq:LB1:`r},
\begin{align*}
  \frac{`r}{x(E')} &= \max_{`l\in `L} \sum_{B\nsupseteq A} `l(B) \frac{x(E(B))}{x(E')}\\
  &= \max_{`l\in `L} \sum_{B\nsupseteq A} `l(B) p(E(B))
\end{align*}
where the last equality is obtained by setting
\begin{align}
  p(e) &= \frac{x_e}{x(E')} && \forall e\in E' \label{eq:p}\\
  p(E'') &=\sum_{e\in E''} p(e) && \forall E''\subseteq E'.\notag
\end{align}
Note that $p$ is a probability distribution over $E'$ since $p(E')=1$ and $p(e)\in [0,1]$ for $e\in E'$ by \eqref{eq:LB1:x}. Let $\rsfsP(E')$ be the set of all possible distributions over $E'$. Then,
\begin{align*}
  \frac{`r}{x(E')} &\utag{b}\geq \min_{P_{\R{e}}\in \rsfsP(E')} \max_{`l\in `L} \sum_{B\nsupseteq A} `l(B) E`1[\ds1\Set{`x(\R{e})\subseteq B}`2]\\
  &\utag{c}= \max_{`l\in `L} \min_{P_{\R{e}}\in \rsfsP(E')}  E`1[\sum_{B\nsupseteq A:`x(\R{e})\subseteq B} `l(B)`2]\\
  &\utag{d}= \max_{`l\in `L}  \max_{e\in E'} \sum_{B\nsupseteq A:`x(e)\subseteq B} `l(B)\\
  &= \max_{`l\in `L}`a(`l)
\end{align*}
which, together with \uref{a}, gives \eqref{eq:LB1} as desired. 
In \uref{b}, $\ds1\Set{`x(\R{e})\subseteq B}$ corresponds to the indicator random variable that is equal to $1$ if and only if the event $`x(\R{e})\subseteq B$ happens, and so the expectation is $P_{\R{e}}(E(B))$ by the definition~\eqref{eq:EB} of $E(B)$. The inequality \uref{b} holds with equality if $P_{\R{e}}=p$ defined in \eqref{eq:p} in terms of $x_{E'}$ is an optimal solution. \uref{c} is by the minimax theorem, since the objective function is linear in both $P_{\R{e}}$ and $`l$ over compact convex sets $\rsfsP(E')$ and $`L$ respectively. \uref{d} is because the expectation over $\R{e}$ is no smaller than the minimum over $e\in E'$. The last equality is by the definition~\eqref{eq:`a} of $`a(`l)$.

Finally, consider proving the equality condition. If $R=0$, the inequalities in \eqref{eq:LB} holds with equality trivially. Similarly, if $`a(`l)\geq 1$, then $\frac1{`a(`l)}-1\leq 0$ and so the inequalities in \eqref{eq:LB} must hold with equality. Consider the remaining case
\begin{align}
  `a(`l) &<1 \forall `l\in `L\label{eq:LB:eq1}\\
  H(\RX_e) &> R> 0 \quad \forall e\in E'. \label{eq:LB:eq2}
\end{align}
We first argue that the constraint $x_e\leq H(\RX_e)$ in~\eqref{eq:LB1:`r} can be removed without changing the maximization in \eqref{eq:LB1}, i.e.,
\begin{align}
  \begin{split}
    \CS(R) - H(\RX_{E`/E'}) &= \max\{ x(E') - `r | `r \leq R\\
    &\kern1em`r = \max_{`l\in `L} \sum_{B\nsupseteq A} `l(B) x(E(B))\\
    &\kern1em x_e\geq 0 \quad \forall e\in E'
  \}.  
  \end{split}
  \label{eq:LB2}
\end{align}
which differs from \eqref{eq:LB1} in the last constraint. Suppose to the contrary that one can have a feasible solution to the above with $x_{e'}>H(\RX_{e'})$ for some $e'\in E'$. Then, with for some partition $\Set{B_1,B_2}$ of $V`/`x(e')$ such that $B_i\nsupseteq A$ for $i\in \Set{1,2}$, define
\begin{align*}
  `l'(B) &=
  \begin{cases}
    1 & B=`x(e'),B_1,B_2\\
    0 & \text{otherwise}
  \end{cases}
\end{align*}
which is a fraction partition in $`L$. It follows from the definition of $`r$ and the constraint $x_e\geq 0$ that
\begin{align*}
  `r &\geq `l'(`x(e')) x_{e'} \geq H(\RX_{e'}) > R
\end{align*}
which is the desired contradiction.

Next, we argue that the constraint~$`r\leq R$ in \eqref{eq:LB2} must be tight for the optimal solution. 
Suppose to the contrary that the constraint is slack. Then, \eqref{eq:LB2} simplifies to
\begin{align}
  \begin{split}
  \CS(R) - H(\RX_{E`/E'}) &= \max_{x_{E'}\in `R_+^{E'}} x(E')\\
    &\kern3em  - \max_{`l\in `L} \sum_{B\nsupseteq A} `l(B) x(E(B))
  \end{split}\label{eq:LB3}\\
    &> 0,\notag  
\end{align}
where the last inequality is because
\begin{align*}
  1 &> \max_{`l\in `L} `a(`l) \\
  &= \min_{P_{\R{e}}\in \rsfsP(E')} \max_{`l\in `L} \sum_{B\nsupseteq A} `l(B) E`1[\ds1\Set{`x(\R{e})\subseteq B}`2]\\
  &= \min_{x_{E'}\in `R_+^{E'}} \max_{`l\in `L} \sum_{B\nsupseteq A} `l(B) \frac{x(E(B))}{x(E')},
\end{align*}
where the first is by \uref{eq:LB:eq1}, the second equality is by \uref{c} and \uref{d}, and the last equality is by rewriting $P_{\R{e}}=p$ defined in \eqref{eq:p}.
Now, given any feasible solution $x_{E'}$ to \eqref{eq:LB3}, $2x_{E'}$ is a strictly better solution, and so the maximum in \eqref{eq:LB3} must be unbounded, contradicting the fact that \eqref{eq:LB1} is bounded by $H(\RX_{E'})$. 

Altogether, for any optimal solution $x_{E'}$ and $`r=R>0$, \uref{a} holds with equality as $`r=R$, and
\begin{align*}
  \frac{`r}{x(E')}
  &= \min_{x_{E'}\in `R_+^{\abs{E'}}} \max_{`l\in `L} \sum_{B\nsupseteq A} `l(B) x(E(B))\\
  &= \max_{`l\in `L} `a(`l),
\end{align*}
i.e., \uref{b} also holds with equality. This completes the proof.

\subsection{Proof of Proposition~\ref{pro:`a}}
\label{sec:proof:`a}
The first inequality in \eqref{eq:slope} follows immediately from \eqref{eq:LB} with any $R>0$. Equality holds trivially if $E'\in E(C_1)\cup E(C_2)$ for some bipartition $\Set{C_1,C_2}$ such that $C_i\cap A\neq `0$ for all $i\in \Set{1,2}$. This is because one can show that $\max_{`l\in `L}`a(`l)=1$ with $`l(B)=1$ for $B=V`/C_i$, $i\in \Set{1,2}$, and $`l(B)=0$ otherwise.

To show the second inequality, let
\begin{align*}
  `l'(B)  &= \frac1{\abs{A}}\ds1\Set{\exists j\in A, B=V`/\Set{j}}\\
        &\quad + \frac1{\abs{A}(\abs{A}-1)}\ds1\Set{\exists j\in A, B=V`/\Set{j}}
\end{align*}
for $B\subseteq V:B\nsupseteq A$. Then, $`l'\in `L$ because, for $i\in V$,
\begin{align*}
  \sum_{B\nsupseteq A:i\in B} `l'(B)
  &= \frac1{\abs{A}}\overbrace{\abs{\Set{V`/\Set{j}\mid j\in A`/\Set{i}}}}^{`(1)} \\
  &\quad + \frac1{\abs{A}(\abs{A}-1)}\underbrace{\abs{\Set{A`/\Set{j}\mid j\in A`/\Set{i}}}}_{`(2)}\\
  &= \begin{cases}
    \frac{\abs{A}-1}{\abs{A}} + \frac1{\abs{A}} & i\in A \quad \begin{aligned}[t]
      \because `(1)&=`(2)\\ &=\abs{A}-1
    \end{aligned}\\
    \frac{\abs{A}}{\abs{A}} + \frac0{\abs{A}} & i\in V`/A \quad \begin{aligned}[t]
      \because `(1)&=\abs{A},\\`(2)&=0
    \end{aligned}
  \end{cases}\\
  &= 1.
\end{align*}
By \eqref{eq:`a},
\begin{align*}
  `a(`l')&=\min_{e\in E'} \sum_{B\nsupseteq A:`x(e)\subseteq B} `l'(B)\\
  &= \min_{e\in E'} \frac1{\abs{A}}\overbrace{\abs{\Set{V`/\Set{j}\mid `x(e)\subseteq V`/\Set{j}}}}^{`(3)} \\
  &\quad + \frac1{\abs{A}(\abs{A}-1)}\underbrace{\abs{\Set{A`/\Set{j}\mid j\in A,`x(e)\subseteq A`/\Set{i}}}}_{`(4)}
\end{align*}
\begin{itemize}
  \item Consider the case $V`/A=`0'$. Then, 
  \begin{align*}
    `(3)&=`(4)=\abs{A`/`x(e)}
  \end{align*}
  and so
  \begin{align*}
    `a(`l') &= \min_{e\in E'} \underbrace{\abs{A`/`x(e)}}_{\geq \abs{A}-d} \overbrace{`1[\frac{1}{\abs{A}} + \frac1{\abs{A}(\abs{A}-1)}`2]}^{=\frac1{\abs{A}-1}} \\
    &\geq \frac{\abs{A}-d}{\abs{A}-1}
  \end{align*}
  \item Consider the remaining case $V`/A\neq`0$. Then,
  \begin{align*}
    `(3)&=\abs{A`/`x(e)}\\
    `(4)&=\begin{cases}
      \abs{A`/`x(e)} & `x(e)\subseteq A\\
      0 & \text{otherwise.}
    \end{cases}
  \end{align*}
  Thus,
  \begin{align*}
    `a(`l') &= \min_{e\in E'} \abs{A`/`x(e)} `1[\frac{1}{\abs{A}} + \frac{\ds1\Set{`x(e)\subseteq A}}{\abs{A}(\abs{A}-1)}`2] \\
    &= \min_{e\in E'} \begin{cases}
      \frac{\abs{A`/`x(e)}}{\abs{A}-1} & `x(e)\subseteq A\\
      \frac{\abs{A`/`x(e)}}{\abs{A}} & \text{otherwise.}
    \end{cases}\\
    &\geq \begin{cases}
      \frac{\max\Set{\abs{A}-d,1}}{\abs{A}-1} & `x(e)\subseteq A\\
      \frac{\max\Set{\abs{A}-d+1,1}}{\abs{A}} & \text{otherwise.}
    \end{cases}\\
    &= \begin{cases}
      \frac{\max\Set{\abs{A}-d,1}}{\abs{A}-1} & d<\abs{A}\\
      \frac{\max\Set{\abs{A}-d+1,1}}{\abs{A}} & \text{otherwise.}
    \end{cases}\\
    &= \begin{cases}
      \frac{\abs{A}-d}{\abs{A}-1} & d<\abs{A}\\
      \frac{1}{\abs{A}} & \text{otherwise.}
    \end{cases}
  \end{align*}
\end{itemize}
Combining the two cases above, we have
\begin{align*}
  `a(`l') &\geq 
  \begin{cases}
    \frac{1}{\abs{A}} & d\geq \abs{A},V`/A\neq `0\\
    \frac{\abs{A}-d}{\abs{A}-1} & \text{otherwise.}
  \end{cases}
\end{align*}
Hence,
\begin{align*}
  \frac1{\max_{`l\in `L} `a(`l)} -1 &\leq \frac1{ `a(`l')} -1\\
  &\leq 
  \begin{cases}
    \abs{A}-1 & d\geq \abs{A},V`/A\neq `0\\
    \frac{d-1}{\abs{A}-d} & \text{otherwise,}
  \end{cases}  
\end{align*}
which can be shown to simplify to the second inequality in \eqref{eq:slope} as desired.
For a complete $d$-uniform hypergraph, all the above inequalities can be satisfied with equality, and so equality can also hold for the second inequality in \eqref{eq:slope}.

\section*{Acknowledgment} 

The author would like to thank Prof.\ Navin Kashyap, Praneeth Kumar Vippathalla, and Qiaoqiao Zhao for their valuable comments and discussions.

\bibliographystyle{IEEEtran}
\bibliography{IEEEabrv,ref}

\end{document}

%% file: preamble.tex
\usepackage[numbers,sort&compress]{natbib}

\makeatletter

\usepackage{etex}

\usepackage{zref-savepos}

\newcounter{mnote}

\def\xmarginnote{%
  \xymarginnote{\hskip -\marginparsep \hskip -\marginparwidth}}

\def\ymarginnote{%
  \xymarginnote{\hskip\columnwidth \hskip\marginparsep}}

\long\def\xymarginnote#1#2{%
\vadjust{#1%
\smash{\hbox{{%
        \hsize\marginparwidth
        \@parboxrestore
        \@marginparreset
\footnotesize #2}}}}}

\def\mnoteson{%
\gdef\mnote##1{\refstepcounter{mnote}\label{##1}%
  \zsavepos{##1}%
  \ifnum20432158>\number\zposx{##1}%
  \xmarginnote{{\color{blue}\bf $\langle$\arabic{mnote}$\rangle$}}%
  \else
  \ymarginnote{{\color{blue}\bf $\langle$\arabic{mnote}$\rangle$}}%
  \fi%
}
  }
\gdef\mnotesoff{\gdef\mnote##1{}}
\mnoteson
\mnotesoff

\usepackage{framed}
\usepackage{subcaption}
\usepackage{comment}
\usepackage[svgnames,dvipsnames]{xcolor}
\usepackage[all]{xy}
\usepackage{tikz}
\usetikzlibrary{positioning,matrix,through,calc,arrows,fit,shapes,decorations.pathreplacing,decorations.markings,}

\tikzstyle{block} = [draw,fill=blue!20,minimum size=2em]

\usepackage{qsymbols,amssymb,mathrsfs}
\usepackage{amsmath}
\usepackage[standard,thmmarks]{ntheorem}
\theoremstyle{plain}
\theoremsymbol{\ensuremath{_\vartriangleleft}}
\theorembodyfont{\itshape}
\theoremheaderfont{\normalfont\bfseries}
\theoremseparator{}

\theoremstyle{nonumberplain}
\theoremheaderfont{\scshape}
\theorembodyfont{\normalfont}
\theoremsymbol{\ensuremath{_\blacktriangleleft}}

\theoremnumbering{arabic}
\theoremstyle{plain}
\usepackage{latexsym}
\theoremsymbol{\ensuremath{_\Box}}
\theorembodyfont{\itshape}
\theoremheaderfont{\normalfont\bfseries}
\theoremseparator{}

\theorembodyfont{\upshape}
\theoremprework{\bigskip\hrule}
\theorempostwork{\hrule\bigskip}

\usepackage[overload]{empheq} 

\let\iftwocolumn\if@twocolumn
\g@addto@macro\@twocolumntrue{\let\iftwocolumn\if@twocolumn}
\g@addto@macro\@twocolumnfalse{\let\iftwocolumn\if@twocolumn}

\mathtoolsset{showonlyrefs=false,showmanualtags}
\let\underbrace\LaTeXunderbrace 
\let\overbrace\LaTeXoverbrace
\renewcommand{\eqref}[1]{\textup{(\refeq{#1})}} 
\newtagform{brackets}[]{(}{)}   
\usetagform{brackets}

\usepackage[Smaller]{cancel}

\PassOptionsToPackage{breaklinks,hyperindex=true,backref=false,bookmarksnumbered,bookmarksopen,linktocpage,colorlinks,linkcolor=BrickRed,citecolor=OliveGreen,urlcolor=Blue,pdfstartview=FitH}{hyperref}
\usepackage{hyperref}

\usepackage{graphicx,psfrag}
\graphicspath{{figure/}{image/}} 

\usepackage{diagbox} 

\usepackage{algorithm2e}
\usepackage{listings} 
\lstdefinelanguage{Maple}{
  morekeywords={proc,module,end, for,from,to,by,while,in,do,od
    ,if,elif,else,then,fi ,use,try,catch,finally}, sensitive,
  morecomment=[l]\#,
  morestring=[b]",morestring=[b]`}[keywords,comments,strings]
\DeclareMathAlphabet{\mathpzc}{OT1}{pzc}{m}{it}
\usepackage{upgreek} 
\usepackage{dsfont}  

\def\multi@nostar#1#2{%
  \expandafter\def\csname multi#1\endcsname##1{%
    \if ##1.\let\next=\relax \else
    \def\next{\csname multi#1\endcsname}     
    \expandafter\newcommand\csname #1##1\endcsname{#2}
    \fi\next}}

\def\multi@star#1#2{%
  \expandafter\def\csname #1\endcsname##1{#2}
  \multi@nostar{#1}{#2}
}

\newcommand{\multi}{%
  \@ifstar \multi@star \multi@nostar}

\multi*{rm}{\mathrm{#1}}
\multi*{mc}{\mathcal{#1}}
\multi*{op}{\mathop {\operator@font #1}}
\multi*{ds}{\mathds{#1}}
\multi*{set}{\mathcal{#1}}
\multi*{rsfs}{\mathscr{#1}}
\multi*{pz}{\mathpzc{#1}}
\multi*{M}{\boldsymbol{#1}}
\multi*{R}{\mathsf{#1}}
\multi*{RM}{\M{\R{#1}}}
\multi*{bb}{\mathbb{#1}}
\multi*{td}{\tilde{#1}}
\multi*{tR}{\tilde{\mathsf{#1}}}
\multi*{trM}{\tilde{\M{\R{#1}}}}
\multi*{tset}{\tilde{\mathcal{#1}}}
\multi*{tM}{\tilde{\M{#1}}}
\multi*{baM}{\bar{\M{#1}}}
\multi*{ol}{\overline{#1}}

\multirm  ABCDEFGHIJKLMNOPQRSTUVWXYZabcdefghijklmnopqrstuvwxyz.
\multiol  ABCDEFGHIJKLMNOPQRSTUVWXYZabcdefghijklmnopqrstuvwxyz.
\multitR   ABCDEFGHIJKLMNOPQRSTUVWXYZabcdefghijklmnopqrstuvwxyz.
\multitd   ABCDEFGHIJKLMNOPQRSTUVWXYZabcdefghijklmnopqrstuvwxyz.
\multitset ABCDEFGHIJKLMNOPQRSTUVWXYZabcdefghijklmnopqrstuvwxyz.
\multitM   ABCDEFGHIJKLMNOPQRSTUVWXYZabcdefghijklmnopqrstuvwxyz.
\multibaM   ABCDEFGHIJKLMNOPQRSTUVWXYZabcdefghijklmnopqrstuvwxyz.
\multitrM   ABCDEFGHIJKLMNOPQRSTUVWXYZabcdefghijklmnopqrstuvwxyz.
\multimc   ABCDEFGHIJKLMNOPQRSTUVWXYZabcdefghijklmnopqrstuvwxyz.
\multiop   ABCDEFGHIJKLMNOPQRSTUVWXYZabcdefghijklmnopqrstuvwxyz.
\multids   ABCDEFGHIJKLMNOPQRSTUVWXYZabcdefghijklmnopqrstuvwxyz.
\multiset  ABCDEFGHIJKLMNOPQRSTUVWXYZabcdefghijklmnopqrstuvwxyz.
\multirsfs ABCDEFGHIJKLMNOPQRSTUVWXYZabcdefghijklmnopqrstuvwxyz.
\multipz   ABCDEFGHIJKLMNOPQRSTUVWXYZabcdefghijklmnopqrstuvwxyz.
\multiM    ABCDEFGHIJKLMNOPQRSTUVWXYZabcdefghijklmnopqrstuvwxyz.
\multiR    ABCDEFGHIJKL NO QR TUVWXYZabcd fghijklmnopqrstuvwxyz.
\multibb   ABCDEFGHIJKLMNOPQRSTUVWXYZabcdefghijklmnopqrstuvwxyz.
\multiRM   ABCDEFGHIJKLMNOPQRSTUVWXYZabcdefghijklmnopqrstuvwxyz.

\newcommand{\dotleq}{\buildrel \textstyle  .\over {\smash{\lower
      .2ex\hbox{\ensuremath\leqslant}}\vphantom{=}}}
\newcommand{\dotgeq}{\buildrel \textstyle  .\over {\smash{\lower
      .2ex\hbox{\ensuremath\geqslant}}\vphantom{=}}}

\newcommand{\bM}{\begin{bmatrix}}
\newcommand{\eM}{\end{bmatrix}}
\newcommand{\bSM}{\left[\begin{smallmatrix}}
\newcommand{\eSM}{\end{smallmatrix}\right]}
\renewcommand*\env@matrix[1][*\c@MaxMatrixCols c]{%
  \hskip -\arraycolsep
  \let\@ifnextchar\new@ifnextchar
  \array{#1}}

\newqsymbol{`N}{\mathbb{N}}
\newqsymbol{`R}{\mathbb{R}}
\newqsymbol{`Z}{\mathbb{Z}}

\usepackage{mleftright}

\newqsymbol{`|}{\mid}
\newqsymbol{`8}{\infty}
\newqsymbol{`1}{\mleft}
\newqsymbol{`2}{\mright}
\newqsymbol{`6}{\partial}
\newqsymbol{`0}{\emptyset}
\newqsymbol{`-}{\leftrightarrow}

\DeclarePairedDelimiter\abs{\lvert}{\rvert}

\DeclarePairedDelimiter\Set{\{}{\}}
 
\newcommand{\imod}[1]{\allowbreak\mkern10mu({\operator@font mod}\,\,#1)}

\newcommand{\threecols}[3]{
\hbox to \textwidth{%
      \normalfont\rlap{\parbox[b]{\textwidth}{\raggedright#1\strut}}%
        \hss\parbox[b]{\textwidth}{\centering#2\strut}\hss
        \llap{\parbox[b]{\textwidth}{\raggedleft#3\strut}}%
    }
}

\newcommand{\reason}[2][\relax]{
  \ifthenelse{\equal{#1}{\relax}}{
    \left(\text{#2}\right)
  }{
    \left(\parbox{#1}{\raggedright #2}\right)
  }
}

\newcommand{\utag}[2]{\mathop{#2}\limits^{\text{(#1)}}}
\newcommand{\uref}[1]{(#1)}

\let\SavedDoubleVert\relax
\let\protect\relax
{\catcode`\|=\active
  \xdef\extendvert{\protect\expandafter\noexpand\csname extendvert \endcsname}
  \expandafter\gdef\csname extendvert \endcsname#1{\mskip-5mu \left.%
      \ifx\SavedDoubleVert\relax \let\SavedDoubleVert\|\fi
     \:{\let\|\SetDoubleVert
       \mathcode`\|32768\let|\SetVert
     #1}\:\right.\mskip-5mu}
}
\def\SetVert{\@ifnextchar|{\|\@gobble}
    {\egroup\;\mid@vertical\;\bgroup}}
\def\SetDoubleVert{\egroup\;\mid@dblvertical\;\bgroup}

\begingroup
 \edef\@tempa{\meaning\middle}
 \edef\@tempb{\string\middle}
\expandafter \endgroup \ifx\@tempa\@tempb
 \def\mid@vertical{\middle|}
 \def\mid@dblvertical{\middle\SavedDoubleVert}
\else
 \def\mid@vertical{\mskip1mu\vrule\mskip1mu}
 \def\mid@dblvertical{\mskip1mu\vrule\mskip2.5mu\vrule\mskip1mu}
\fi

\makeatother

\usepackage{fouridx}
\usepackage{framed}
\usetikzlibrary{positioning,matrix}

\usepackage{paralist}
\usepackage{enumerate}

\usepackage[normalem]{ulem}

{\endMakeFramed}

\newenvironment{ybox}{
	\setlength{\FrameSep}{1.5mm}
	\setlength{\FrameRule}{0mm}
  \MakeFramed {\FrameRestore}}%
{\endMakeFramed}

\newenvironment{gbox}{
	\setlength{\FrameSep}{1.5mm}
\setlength{\FrameRule}{0mm}
  \MakeFramed {\FrameRestore}}%
{\endMakeFramed}

{\endMakeFramed}

 {\endMakeFramed}

\usepackage{enumitem}

\usepackage{etoolbox}

\let\theparentequation\theequation
\patchcmd{\theparentequation}{equation}{parentequation}{}{}

\renewenvironment{subequations}[1][]{
	\refstepcounter{equation}%
	\setcounter{parentequation}{\value{equation}}
	\setcounter{equation}{0}
	\def\theequation{\theparentequation\alph{equation}}%
	\let\parentlabel\label
	\ifx\\#1\\\relax\else\label{#1}\fi
	\ignorespaces
}{%
	\setcounter{equation}{\value{parentequation}}
	\ignorespacesafterend
}

\newcommand*{\nextParentEquation}[1][]{
	\refstepcounter{parentequation}
	\setcounter{equation}{0}
	\ifx\\#1\\\relax\else\parentlabel{#1}\fi
}

%% file: main.bbl
\begin{thebibliography}{10}
\providecommand{\url}[1]{#1}
\csname url@samestyle\endcsname
\providecommand{\newblock}{\relax}
\providecommand{\bibinfo}[2]{#2}
\providecommand{\BIBentrySTDinterwordspacing}{\spaceskip=0pt\relax}
\providecommand{\BIBentryALTinterwordstretchfactor}{4}
\providecommand{\BIBentryALTinterwordspacing}{\spaceskip=\fontdimen2\font plus
\BIBentryALTinterwordstretchfactor\fontdimen3\font minus
  \fontdimen4\font\relax}
\providecommand{\BIBforeignlanguage}[2]{{%
\expandafter\ifx\csname l@#1\endcsname\relax
\typeout{** WARNING: IEEEtran.bst: No hyphenation pattern has been}%
\typeout{** loaded for the language `#1'. Using the pattern for}%
\typeout{** the default language instead.}%
\else
\language=\csname l@#1\endcsname
\fi
#2}}
\providecommand{\BIBdecl}{\relax}
\BIBdecl

\bibitem{ahlswede93}
R.~Ahlswede and I.~Csisz{\'{a}}r, ``Common randomness in information theory and
  cryptography---{P}art {I}: Secret sharing,'' \emph{IEEE Transactions on
  Information Theory}, vol.~39, no.~4, pp. 1121--1132, Jul. 1993.

\bibitem{maurer93}
U.~M. Maurer, ``Secret key agreement by public discussion from common
  information,'' \emph{IEEE Transactions on Information Theory}, vol.~39,
  no.~3, pp. 733--742, 1993.

\bibitem{bennett1988privacy}
C.~H. Bennett, G.~Brassard, and J.-M. Robert, ``Privacy amplification by public
  discussion,'' \emph{SIAM journal on Computing}, vol.~17, no.~2, pp. 210--229,
  1988.

\bibitem{csiszar00}
I.~Csisz{\'a}r and P.~Narayan, ``Common randomness and secret key generation
  with a helper,'' \emph{IEEE Transactions on Information Theory}, vol.~46,
  no.~2, pp. 344--366, 2000.

\bibitem{csiszar04}
I.~Csisz{\'{a}}r and P.~Narayan, ``Secrecy capacities for multiple terminals,''
  \emph{IEEE Transactions on Information Theory}, vol.~50, no.~12, pp.
  3047--3061, Dec. 2004.

\bibitem{tyagi13}
H.~Tyagi, ``Common information and secret key capacity,'' \emph{IEEE
  Transactions on Information Theory}, vol.~59, no.~9, pp. 5627--5640, Sep.
  2013.

\bibitem{LCV16}
J.~Liu, P.~Cuff, and S.~Verd\'{u}, ``Secret key generation with limited
  interaction,'' \emph{IEEE Transactions on Information Theory}, vol.~63, pp.
  7358--7381, 2017.

\bibitem{chan18isit}
C.~{Chan}, M.~{Mukherjee}, N.~{Kashyap}, and Q.~{Zhou}, ``Multiterminal secret
  key agreement at asymptotically zero discussion rate,'' in \emph{2018 IEEE
  International Symposium on Information Theory (ISIT)}, June 2018, pp.
  2654--2658.

\bibitem{chan19cs0}
C.~Chan, M.~Mukherjee, P.~K. Vippathalla, and Q.~Zhou, ``Multiterminal secret
  key agreement with nearly no discussion,'' \emph{arXiv preprint
  arXiv:1904.11383}, 2019.

\bibitem{chan19lamination}
C.~{Chan}, M.~{Mukherjee}, N.~{Kashyap}, and Q.~{Zhou}, ``Upper bounds via
  lamination on the constrained secrecy capacity of hypergraphical sources,''
  \emph{IEEE Transactions on Information Theory}, pp. 1--1, 2019.

\bibitem{nitinawarat10}
S.~Nitinawarat and P.~Narayan, ``Perfect omniscience, perfect secrecy, and
  {S}teiner tree packing,'' \emph{IEEE Transactions on Information Theory},
  vol.~56, no.~12, pp. 6490--6500, Dec. 2010.

\bibitem{chan17isit}
C.~Chan, M.~Mukherjee, N.~Kashyap, and Q.~Zhou, ``Secret key agreement under
  discussion rate constraints,'' in \emph{IEEE International Symposium on
  Information Theory Proceedings (ISIT)}, June 2017, pp. 1519--1523.

\bibitem{zhou2018secrecy}
\BIBentryALTinterwordspacing
Q.~Zhou and C.~Chan, ``Secrecy capacity under limited discussion rate for
  minimally connected hypergraphical sources,'' \emph{CoRR}, vol.
  abs/1805.03110, 2018. [Online]. Available:
  \url{http://arxiv.org/abs/1805.03110}
\BIBentrySTDinterwordspacing

\bibitem{zhou2018isit}
------, ``Secrecy capacity under limited discussion rate for minimally
  connected hypergraphical sources,'' in \emph{2018 IEEE International
  Symposium on Information Theory (ISIT)}.\hskip 1em plus 0.5em minus
  0.4em\relax IEEE, 2018, pp. 2664--2668.

\bibitem{MKS16}
M.~Mukherjee, N.~Kashyap, and Y.~Sankarasubramaniam, ``On the public
  communication needed to achieve sk capacity in the multiterminal source
  model,'' \emph{IEEE Transactions on Information Theory}, vol.~62, no.~7, pp.
  3811--3830, July 2016.

\bibitem{chan16isit}
C.~Chan, A.~Al-Bashabsheh, and Q.~Zhou, ``Incremental and decremental secret
  key agreement,'' in \emph{IEEE International Symposium on Information Theory
  Proceedings (ISIT)}, July 2016, pp. 2514--2518.

\bibitem{mukherjee16}
M.~Mukherjee, C.~Chan, N.~Kashyap, and Q.~Zhou, ``Bounds on the communication
  rate needed to achieve {SK} capacity in the hypergraphical source model,'' in
  \emph{IEEE International Symposium on Information Theory Proceedings (ISIT)},
  July 2016, pp. 2504--2508.

\bibitem{chan16itw}
C.~Chan, M.~Mukherjee, N.~Kashyap, and Q.~Zhou, ``When is omniscience a
  rate-optimal strategy for achieving secret key capacity?'' in \emph{IEEE
  Information Theory Workshop (ITW)}, Sep. 2016, pp. 354--358.

\bibitem{chan17oo}
------, ``On the optimality of secret key agreement via omniscience,''
  \emph{IEEE Transactions on Information Theory}, vol.~64, pp. 2371--2389,
  2018.

\bibitem{chan19plska}
C.~{Chan}, N.~{Kashyap}, P.~K. {Vippathalla}, and Q.~{Zhou}, ``One-shot perfect
  secret key agreement for finite linear sources,'' in \emph{2019 IEEE
  International Symposium on Information Theory (ISIT)}, July 2019, pp.
  947--951.

\bibitem{chan17cska}
C.~Chan, ``Compressed secret key agreement:maximizing multivariate mutual
  information per bit,'' \emph{Entropy}, vol.~19, no.~10, 2017.

\bibitem{nitinawarat-ye10}
S.~Nitinawarat, C.~Ye, A.~Barg, P.~Narayan, and A.~Reznik, ``Secret key
  generation for a pairwise independent network model,'' \emph{IEEE
  Transactions on Information Theory}, vol.~56, no.~12, pp. 6482--6489, Dec
  2010.

\bibitem{chan10phd}
C.~Chan, ``Generating secret in a network,'' Ph.D. dissertation, Massachusetts
  Institute of Technology, 2010.

\bibitem{chan11itw}
------, ``Linear perfect secret key agreement,'' in \emph{Information Theory
  Workshop (ITW), 2011 IEEE}.\hskip 1em plus 0.5em minus 0.4em\relax IEEE,
  2011, pp. 723--726.

\bibitem{chan11delay}
------, ``Delay of linear perfect secret key agreement,'' in \emph{Forty-Ninth
  Annual Allerton Conference on Communication, Control, and Computing}, Sep.
  2011.

\bibitem{chan11isit}
------, ``The hidden flow of information,'' in \emph{IEEE International
  Symposium on Information Theory Proceedings (ISIT)}, Jul. 2011.

\bibitem{chan12ud}
------, ``Matroidal undirected network,'' in \emph{IEEE International Symposium
  on Information Theory Proceedings (ISIT)}, July 2012, pp. 1498--1502.

\bibitem{courtade16}
T.~A. Courtade and T.~R. Halford, ``Coded cooperative data exchange for a
  secret key,'' \emph{IEEE Transactions on Information Theory}, vol.~62, no.~7,
  pp. 3785--3795, July 2016.

\bibitem{chan10md}
C.~Chan and L.~Zheng, ``Mutual dependence for secret key agreement,'' in
  \emph{Proceedings of 44th Annual Conference on Information Sciences and
  Systems}, 2010.

\bibitem{csiszar81}
I.~Csisz{\'{a}}r and J.~K{\"{o}}rner, \emph{Information Theory: Coding Theorems
  for Discrete Memoryless Systems}.\hskip 1em plus 0.5em minus 0.4em\relax
  Akad{\'{e}}miai Kiad{\'{o}}, Budapest, 1981.

\bibitem{schrijver02}
A.~Schrijver, \emph{Combinatorial Optimization: Polyhedra and
  Efficiency}.\hskip 1em plus 0.5em minus 0.4em\relax Springer, 2002.

\bibitem{chan17ooa}
\BIBentryALTinterwordspacing
C.~Chan, M.~Mukherjee, N.~Kashyap, and Q.~Zhou, ``On the optimality of secret
  key agreement via omniscience,'' \emph{CoRR}, vol. abs/1702.07429, 2017.
  [Online]. Available: \url{http://arxiv.org/abs/1702.07429}
\BIBentrySTDinterwordspacing

\bibitem{csiszar08}
I.~Csisz{\'{a}}r and P.~Narayan, ``Secrecy capacities for multiterminal channel
  models,'' \emph{IEEE Transactions on Information Theory}, vol.~54, no.~6, pp.
  2437--2452, June 2008.

\bibitem{chan2008tightness}
C.~Chan, ``On tightness of mutual dependence upperbound for secret-key capacity
  of multiple terminals,'' \emph{arXiv preprint arXiv:0805.3200}, 2008.

\bibitem{chan15mi}
C.~Chan, A.~Al-Bashabsheh, J.~Ebrahimi, T.~Kaced, and T.~Liu, ``Multivariate
  mutual information inspired by secret-key agreement,'' \emph{Proceedings of
  the IEEE}, vol. 103, no.~10, pp. 1883--1913, Oct 2015.

\bibitem{chan16cluster}
C.~Chan, A.~Al-Bashabsheh, Q.~Zhou, T.~Kaced, and T.~Liu, ``Info-clustering: A
  mathematical theory for data clustering,'' \emph{IEEE Transactions on
  Molecular, Biological and Multi-Scale Communications}, vol.~2, no.~1, pp.
  64--91, June 2016.

\bibitem{chan17idska}
C.~Chan, A.~Al-Bashabsheh, and Q.~Zhou, ``Change of multivariate mutual
  information: From local to global,'' \emph{IEEE Transactions on Information
  Theory}, vol.~PP, no.~99, pp. 1--1, 2017.

\bibitem{chan13itw}
C.~Chan, K.~W. Shum, and Q.~T. Sun, ``Combinatorial flow over cyclic linear
  networks,'' in \emph{IEEE Information Theory Workshop (ITW)}, Sep. 2013, pp.
  1--5.

\bibitem{chan13isit}
C.~Chan, ``Cyclic linking network,'' in \emph{IEEE International Symposium on
  Information Theory Proceedings (ISIT)}, July 2013, pp. 789--793.

\bibitem{csiszar2011information}
I.~Csiszar and J.~K{\"o}rner, \emph{Information theory: coding theorems for
  discrete memoryless systems}, 2nd~ed.\hskip 1em plus 0.5em minus 0.4em\relax
  Cambridge University Press, 2011.

\bibitem{fujishige78}
S.~Fujishige, ``Polymatroidal dependence structure of a set of random
  variables,'' \emph{Information and Control}, vol.~39, no.~1, pp. 55 -- 72,
  1978.

\bibitem{fujishige05}
------, \emph{Submodular functions and optimization}, 2nd~ed.\hskip 1em plus
  0.5em minus 0.4em\relax Elsevier, 2005.

\end{thebibliography}
